\documentclass[aps,prd,showpacs,floatfix,
onecolumn,
groupedaddress,
nofootinbib,preprintnumbers,
10pt]{revtex4-2}
\makeatletter
\def\p@subsection{}
\makeatother

\usepackage{graphicx,amsmath,amssymb,lipsum,bm}

\usepackage[dvipsnames]{xcolor}
\definecolor{xlinkcolor}{rgb}{0.7752941176470588, 0.22078431372549023, 0.2262745098039215}
\definecolor{BrickRed}{rgb}{0.7752941176470588, 0.22078431372549023, 0.2262745098039215}
\definecolor{xlinkcolor}{HTML}{1c1e94}
\usepackage{booktabs} 
\usepackage[normalem]{ulem}
\usepackage[utf8]{inputenc} 
\usepackage{mathtools}
\usepackage{aux_files/aas_macro}
\usepackage{xspace}

\usepackage{ulem}
\usepackage{diagbox}

\usepackage[dvipsnames]{xcolor}
\usepackage{mathrsfs}
\usepackage{tabularx} 

\usepackage[colorlinks=true,citecolor=xlinkcolor,linkcolor=xlinkcolor,urlcolor=xlinkcolor, backref=false,pdfborder={0 0 0}]{hyperref}
\newcommand{\beqa}{\begin{eqnarray}}
\newcommand{\eeqa}{\end{eqnarray}}

\newcommand{\be}{\begin{equation}}
\newcommand{\ee}{\end{equation}}
\newcommand{\beq}{\begin{equation}}
\newcommand{\eeq}{\end{equation}}

\renewcommand\k{{\bm k}}

\newcommand{\bseq}{\begin{subequations}}
\newcommand{\eseq}{\end{subequations}}

\renewcommand{\ln}{\mathop{\rm ln}\nolimits}

\def\ltsima{$\; \buildrel < \over \sim \;$\xspace}
\def\gtsima{$\; \buildrel > \over \sim \;$\xspace}
\def\simlt{\lower.5ex\hbox{\ltsima}}
\def\simgt{\lower.5ex\hbox{\gtsima}}

\newcommand{\hMpcinv}{\,h\, {\rm Mpc}^{-1}\xspace}

\newcommand{\Mpch}{\, h^{-1}\mathrm{Mpc}\, }

\newcommand{\kmax}{\, k_{\rm max}\, }
\newcommand{\knl}{\, k_{\rm NL}\, }

\newcommand{\Lya}{Ly-$\alpha$\xspace}

\newcommand{\kpar}{k_{\parallel}}

\def\gsim{\raise0.3ex\hbox{$\;>$\kern-0.75em\raise-1.1ex\hbox{$\sim\;$}}}
\def\lsim{\raise0.3ex\hbox{$\;<$\kern-0.75em\raise-1.1ex\hbox{$\sim\;$}}}

\def\beqn#1{\begin{equation}\label{#1}}
\def\eeqn{\end{equation}}

\def\beqa#1{\begin{eqnarray}\label{#1}}
\def\eeqa{\end{eqnarray}}

\newcommand{\abacus}{\textsc{AbacusSummit}\xspace}

\def\kmax{{k_\text{max}}}

\def\Mpch{h^{-1}{\text{Mpc}}}

\def\Z2{$\mathcal{Z_2}$}

\newcommand {\ignore}[1]{}

\renewcommand{\arraystretch}{1.3} 

\DeclareRobustCommand{\ion}[2]{%
\relax\ifmmode
\ifx\testbx\f@series
{\mathbf{#1\,\mathsc{#2}}}\else
{\mathrm{#1\,\mathsc{#2}}}\fi
\else\textup{#1\,{\mdseries\textsc{#2}}}%
\fi}

\newcommand{\MIT}{Center for Theoretical Physics -- a Leinweber Institute, Massachusetts Institute of Technology, Cambridge, MA 02139, USA}
\newcommand{\IAIFI}{The NSF AI Institute for Artificial Intelligence and Fundamental Interactions, Cambridge, MA 02139, USA}

\begin{document}
\newcommand{\sig}{\sigma_8}
\newcommand{\lnAs}{\ln{10^{10} A_s}}
\newcommand{\qpar}{q_{\parallel}}
\newcommand{\qperp}{q_{\perp}}

\preprint{MIT-CTP/6145}

\title{Full-shape analysis of the Lyman-alpha forest power spectrum and bispectrum: validation on 
simulations}

\author{Roger de Belsunce}
\email{belsunce@mit.edu}
\affiliation{\MIT}
\affiliation{\IAIFI}
\author{James M. Sullivan}
\email{jms3@mit.edu}
\thanks{Brinson Prize Fellow}
\affiliation{\MIT}
\affiliation{\IAIFI}
\author{Mikhail M. Ivanov}
\email{ivanov99@mit.edu}
\affiliation{\MIT}
\affiliation{\IAIFI}

\date{\today}

\begin{abstract}
The Lyman-$\alpha$ (\Lya) forest of neutral hydrogen 
provides a large-volume, high-resolution map of cosmological fluctuations from optical spectra at high redshift.
To extract more information from the currently observing Dark Energy Spectroscopic Instrument (DESI), we derive an effective field theory (EFT) model for joint full-shape analyses of the \Lya forest data. 
It combines the one-loop power spectrum of the \Lya forest ($FF$), and of its cross-correlation with dark-matter halos, used as proxies for high-redshift quasars ($FQ$), 
with the tree-level bispectrum of the \Lya forest ($FFF$). 
Importantly, the bispectrum breaks the \Lya-specific degeneracy between the mean flux and the amplitude of the power spectrum. 
We validate our theory framework on idealized large-volume \Lya forest mocks painted onto the \textsc{AbacusSummit} $N$-body simulation suite at redshift $z=2.5$. 
Fitting scales up to $k_\mathrm{max}^{FF\,(FQ)}=0.6\,(0.3)\,h\,\mathrm{Mpc}^{-1}$ in the power spectrum and $k_\mathrm{max}^{FFF}=0.1\,h\,\mathrm{Mpc}^{-1}$ in the bispectrum, our full-shape analysis yields unbiased constraints on the cosmological parameters in simulations. 
For a fully sampled flux field in a simulated volume of $V=8\,h^{-3}\,{\rm Gpc}^3$, we measure the Hubble constant, $H_0$, to $0.6\%$, the matter density, $\Omega_{\rm m}$, to $1.6\%$ and the amplitude of matter clustering, $\sig$, to $3.5\%$.
Adding the bispectrum improves the constraints on $\sig$ by $28\%$. 
We also demonstrate that the inclusion of the quasar one-loop power spectrum ($QQ$) and tree-level bispectrum ($QQQ$) further improves cosmological constraints by $\approx 35\%$ using $k_\mathrm{max}^{QQ,\,(QQQ)}=0.2,\,(0.08)\,h\,\mathrm{Mpc}^{-1}$.
Our theory framework enables full-shape analyses of high-redshift data from DESI and its successor DESI-II, enhancing knowledge of the growth of structure 
of the Universe during matter-domination. 
\end{abstract}

 \maketitle

\section{Introduction \label{sec:intro}}
Over the last decades, observations of the 
large-scale structure (LSS) of the Universe have provided increasingly precise constraints on the growth of structure and expansion history of our Universe~\cite{DESI:2024mwx, eBOSS:2020yzd}. The large influx of data from  the currently observing Stage-IV Dark Energy Spectroscopic Instrument (DESI;~\cite{DESI:2016}) maps the three-dimensional cosmic web of galaxies, quasars and the Lyman-$\alpha$ (\Lya) forest. Together, these tracers probe an unprecedented volume spanning more than three orders of magnitude in length scale covering a wide redshift range ($0 \simlt z \simlt 4$). Measurements by \textit{Euclid}~\cite{Laureijs:2011gra}, the Nancy Grace Roman Space Telescope~\cite{Spergel:2015sza}, and next-generation Stage-V spectroscopic surveys~\cite{Schlegel:22_spec_roadmap} will further strengthen this observational program.

The \Lya forest, arising from absorption by neutral hydrogen in the intergalactic medium,  maps cosmological perturbations through the lower-density, unvirialized regions of the cosmic web~\cite{McQuinn:2016}. The \Lya forest is therefore highly complementary to galaxies, which trace overdense dark-matter halos. At $2 \simlt z \simlt 4$ it accesses a cosmological volume that will remain beyond the reach of galaxy surveys until Stage-V spectroscopy~\cite{Schlegel:22_spec_roadmap, Schlegel2022MegaMapper, Besuner25:spec_s5, Poppett2026SpecS5}. Its dense sampling along the line of sight further probes matter clustering down to Mpc scales and below. In addition, the amplitude of the intrinsic shot noise of the forest is much weaker than that of galaxies~\cite{Ivanov:2023yla,deBelsunce:2024rvv}, which opens up opportunities
to access more modes in the cosmic variance limit compared to galaxies.

The rapidly expanding DESI \Lya sample, together with advances in three-dimensional measurements~\cite{Font-Ribera:2018, deBelsunce:2024knf, Horowitz:2024nny} and theoretical modeling~\cite{Ivanov:2023yla,deBelsunce:2025bqc,deBelsunceSullivanMcDonald2025}, is making the \Lya forest an increasingly precise LSS tracer. 
On large scales, its BAO feature uniquely constrains the expansion history in the matter-dominated era~\cite{McDonald:2007,Slosar2013,Busca:2013,dMdB:2020,DESI_lya_2024,DESI_lya_dr2}, 
potentially shedding light on the recent distance disagreements with the CMB that hint at evolving dark energy~\cite{Aghanim:2018eyx,DESI2025DR2Cosmo},
negative spatial curvature~\cite{Chen:2025mlf}, a higher optical depth to reionization~\cite{Sailer:2025lxj,Jhaveri:2025neg,Sullivan2026HighTau}, and (more exotically) models with an interacting dark sector~\cite{Khoury:2026svx}. 
Simultaneously, measurements of the broadband three-dimensional correlation function of the forest probe the shape of large-scale \Lya clustering~\cite{Slosar2013,Cuceu:2021hlk,Cuceu:2023,Gordon:2023,Cuceu:2025nvl}, informing the growth of structure. 
On smaller scales, the one-dimensional flux power spectrum, $P_{\rm 1D}$~\cite{Seljak:2005,Viel:2005,McDonald06,PYB13,Chabanier:2019,Pedersen:2020,2023MNRAS.526.5118R,2024MNRAS.tmp..176K}, constrains neutrino masses~\cite{Seljak:2005,Viel:2010,PYB13,Palanque2020}, primordial black holes~\cite{Afshordi:2003,Murgia:2019,Ivanov:2025pbu}, warm-dark-matter models~\cite{Viel:2013,Baur:2016,Irsic17,Kobayashi:2017,Armengaud:2017,Murgia:2018,Garzilli:2019,Irsic:2020,Rogers:2022,Villasenor:2023,Irsic:2023}, early-dark-energy models~\cite{2023PhRvL.131t1001G}, and the thermal state of the ionized IGM~\cite{Zaldarriaga:2002,Meiksin:2009,McQuinn:2016,Viel:2006,Walther:2019,Bolton:2008,Garzilli:2012,Gaikwad:2019,Boera:2019,Gaikwad:2021,Wilson:2022,Villasenor:2022}.

The effective field theory (EFT) of large-scale structure provides a systematic framework 
to realize the potential to constrain cosmological physics with the \Lya forest,
organizing the large-scale dynamics according to the symmetries of the tracer~\cite{Baumann:2010tm,Carrasco:2013mua,Ivanov:2022mrd}. 
For the \Lya forest, these are the equivalence principle and rotations about the line of sight, described by the $SO(2)$ group~\cite{McDonald:2009dh,Givans:2020sez, Desjacques:2018pfv,Chen:2021rnb,Ivanov:2023yla}. 
The general nature of the EFT model for the \Lya forest allows us to 
go beyond state-of-the-art methods for modeling the Lyman-$\alpha$ forest by including the non-Gaussian information of the \Lya flux field.
As a model at the level of cosmological fields, EFT predicts not just the power spectrum\footnote{The EFT \Lya flux power spectrum allows for consistent modeling of the Alcock-Paczy\'nski (AP) effect \cite{Gerardi:2022ncj,Cuceu:2021hlk,DESI:2026lnd,Alcock:1979mp} and mitigates nonlinear biases in the BAO feature~\cite{Hadzhiyska:2025cvk,deBelsunce:2024rvv,Chen:2024tfp}.} \cite{Ivanov:2023yla}, as do current heuristic fitting functions obtained from simulations \cite{Arinyo-i-Prats:2015vqa, Givans:2022qgb}, but also the bispectrum \cite{deBelsunceSullivanMcDonald2025}.
The flux bispectrum measures the same free parameters in the bias model that enter the flux power spectrum. When both statistics are considered jointly this breaks degeneracies and unlocks improved constraints on cosmological parameters.
While the practical estimation of the bispectrum from observational data requires further methodological developments\footnote{A realistic survey geometry requires accounting for the pathological \Lya window function -- skew spectra provide a tractable route to address this challenge~\cite{deBelsunceSullivanMcDonald2025}.}, we illustrate here that the EFT three-dimensional bispectrum contains significant information for cosmological analysis.

In this work, we develop and implement a unified EFT model for the one-loop \Lya forest power spectrum and tree-level bispectrum. We quantify its ability to recover unbiased cosmological constraints from both the full-shape and the BAO feature using large-volume, idealized \Lya forest mocks painted onto the \textsc{AbacusSummit} $N$-body simulations~\cite{Hadzhiyska:2023}. This provides the foundation for applying the joint model to current DESI data, its successor DESI-II, and to future surveys such as the WEAVE-QSO survey~\cite{2016sf2a.conf..259P}, the Prime Focus Spectrograph (PFS;~\cite{2022PFSGE}), and 4MOST~\cite{2019Msngr.175....3D}.

This paper is organized as follows: In Sec.~\ref{sec:mocks} we introduce the used synthetic \Lya forest data. In Sec.~\ref{sec:bk_estimator} we briefly summarize the bispectrum estimator on periodic boxes and in Sec.~\ref{sec:theory} we introduce the bias expansions of the \Lya forest to construct the  one-loop power spectrum and tree-level bispectrum in EFT. In Sec.~\ref{sec:like} we discuss our Gaussian likelihood and prior choices. In Sec.~\ref{sec:results} we present the results of fits to the mock data and conclude in Sec.~\ref{sec:conclusion}.

\section{Simulations}
\label{sec:mocks}

We use the \Lya forest mocks built on the large-volume \textsc{AbacusSummit} suite of $N$-body simulations~\cite{2021MNRAS.508.4017M,Hadzhiyska:2023}.\footnote{The \Lya forest mocks are publicly available on Globus through NERSC SHARE at this link: \url{https://app.globus.org/file-manager?origin_id=9ce29982-eed1-11ed-9bb4-c9bb788c490e&path=\%2F} under the name ``AbacusSummit Lyman Alpha Forest''.} We use a cosmology centered on \emph{Planck} 2018~\cite{Aghanim:2018eyx} with six realizations and two lines of sight. Each mock consists of $6912^3$ particles in a periodic box of side length $L=2000\,\Mpch$, and we use the snapshots at $z=2.5$. The transmitted flux is computed from the dark matter field with the fluctuating Gunn--Peterson approximation (FGPA;~\cite{Croft98}), and the peculiar velocities move the absorption into redshift space. 
Small-scale fluctuations are added such that the one-dimensional flux power spectrum agrees with that of a reference hydrodynamic simulation, and the free parameters of the FGPA are calibrated on IllustrisTNG~\cite{2010MNRAS.401..791S,Weinberger:2019tbd}. Four different calibrations have been provided resulting in Models 1--4 with different bias parameters. In particular, Models 3 and 4 have a larger redshift-space distortion parameter than Models 1 and 2. The construction of the mocks is described in detail in~\cite{Hadzhiyska:2023}. In the present analysis, we will focus on Model 1, both lines of sight ($\hat y$ and $\hat z$) and six realizations for each making a total of 12 mocks as well as the mean data vectors obtained from each of the models. As a variation of our analysis we will also analyze mocks using Model 3.

\section{Measurements}
\label{sec:bk_estimator}

From the transmitted flux $F$ on the simulation mesh we compute the flux density contrast $\delta_F=F/\bar F-1$, where $\bar F$ is the mean flux of the box. The transmitted flux fraction is defined as  $F(x)=\exp{(-\tau(x))}$ where $\tau$ is the optical depth. To reduce the size of the mesh, we keep Fourier modes below the Nyquist frequency by band-pass filtering the high-resolution mesh~\cite[see,][for more details]{Hadzhiyska:2023} but we do not apply a window selection function, i.e.,~draw skewers from the box. The quantities of interest in the present work are  the three-dimensional flux power spectrum, $P_{\rm 3D}(k,\mu)$ (or $P(k,\mu)$ in the following), and multipoles of the flux bispectrum, $B_\ell(k_1,k_2,k_3)$. 

We compute the power spectrum as the average of the squared norm of the Fourier transform of the flux density contrast in bins of the Fourier wavenumber $k$ with the cosine of the angle to the line-of-sight, $\mu=\kpar/k$. It is measured in ten bins in $\mu$ of width $\Delta\mu=0.1$ and in bins of width $\Delta k=0.005\hMpcinv$ which are discussed in detail in Refs.~\cite{Hadzhiyska:2025cvk, Hadzhiyska:2026thl}. 

We measure the bispectrum multipoles using the estimator of Ref.~\cite{Scoccimarro2015:bk_multipoles}.
We order the sides of each triangle as $k_1\geq k_2\geq k_3$ and expand in the cosine of the angle between the longest side and the line of sight, $\mu_1=\hat{\mathbf{q}}_1\cdot\hat{\mathbf{n}}$. In a periodic box of volume $V=L^3$ the estimator reads
\begin{equation}
\label{eq:Bestimator}
\hat B_\ell(k_1,k_2,k_3)=\frac{2\ell+1}{V\,N_T}\sum_{\mathbf{q}_1\in k_1}\sum_{\mathbf{q}_2\in k_2}\sum_{\mathbf{q}_3\in k_3}\delta_K(\mathbf{q}_1+\mathbf{q}_2+\mathbf{q}_3)\,\delta_F(\mathbf{q}_1)\,\delta_F(\mathbf{q}_2)\,\delta_F(\mathbf{q}_3)\,\mathcal{L}_\ell(\mu_1)\,,
\end{equation}
where $\mathbf{q}\in k_i$ denotes the wavevectors of the box in the shell $k_i-\Delta k/2\leq|\mathbf{q}|<k_i+\Delta k/2$, $\delta_K$ is the Kronecker delta, $\mathcal{L}_\ell$ is the Legendre polynomial of order $\ell$, and $\delta_F(\mathbf{q})=(V/N_c)\sum_{\mathbf{x}}e^{-i\mathbf{q}\cdot\mathbf{x}}\delta_F(\mathbf{x})$ is the discrete Fourier transform on a mesh with $N_c$ cells. The normalization
\begin{equation}
\label{eq:NT}
N_T(k_1,k_2,k_3)=\sum_{\mathbf{q}_1\in k_1}\sum_{\mathbf{q}_2\in k_2}\sum_{\mathbf{q}_3\in k_3}\delta_K(\mathbf{q}_1+\mathbf{q}_2+\mathbf{q}_3)\,,
\end{equation}
is the number of closed triangles in the bin. Writing the Kronecker delta as a sum over the mesh, the triple sum becomes a sum over the mesh of the product of three fields. Each of them is the inverse fast Fourier transform (FFT) of $\delta_F$ restricted to one shell, with the weight $\mathcal{L}_\ell(\hat{\mathbf{q}}\cdot\hat{\mathbf{n}})$ for the longest side. $N_T$ is computed in the same way with $\delta_F(\mathbf{q})$ set to one.

The expectation value of Eq.~\eqref{eq:Bestimator} is the bispectrum averaged over the closed triangles of the bin,
\begin{equation}
\label{eq:Bbinned}
B_\ell^{\rm th}(k_1,k_2,k_3)=\frac{2\ell+1}{N_T}\sum_{\mathbf{q}_1\in k_1}\sum_{\mathbf{q}_2\in k_2}\sum_{\mathbf{q}_3\in k_3}\delta_K(\mathbf{q}_1+\mathbf{q}_2+\mathbf{q}_3)\,B(\mathbf{q}_1,\mathbf{q}_2,\mathbf{q}_3)\,\mathcal{L}_\ell(\mu_1)\,,
\end{equation}
which is how we compare the theory model with the measurements. We evaluate this sum exactly for bins with up to $10^7$ triangles. For larger bins whose shortest side has at most $10^3$ modes (squeezed bins), we sum exactly over the modes of the shortest side and take the continuum limit for the other two sides.

We use the implementation of this estimator in \texttt{jax-power},\footnote{Publicly available at \url{https://github.com/adematti/jax-power}.} 
and measure $\ell=0,2$ in bins of width $\Delta k=0.01\hMpcinv$ from $k_{\min}=0.01\hMpcinv$ to $k_{\max}=0.1\hMpcinv$ for our baseline analysis. We additionally investigate two variants with  $k_{\max}=0.06\hMpcinv$ and $k_{\max}=0.15\hMpcinv$. 
For the bispectrum we reduce the field to a $256^3$ mesh. Its Nyquist frequency, $k_{\rm Nyq}=0.40\hMpcinv$, is larger than $1.5\,k_{\max}$. 

\section{Perturbative 2- \& 3-point statistics of the Lyman-Alpha forest \label{sec:model}}

\label{sec:theory}
In this work, we derive and implement the bias expansion  within the EFT framework of the \Lya forest flux decrement to compute the one-loop power spectrum in Sec.~\ref{sec:p1loop} and the tree-level bispectrum in Sec.~\ref{sec:btree}. The EFT has been a remarkably successful program to describe statistics on large scales only using the relevant symmetries, dimensional analysis, the notion of naturalness and the equivalence principle~\cite{ McDonald:2009dh,  Baumann:2010tm, Carrasco:2012cv, Desjacques:2018pfv}. In the context of the \Lya forest early perturbative approaches~\cite{2012JCAP...03..004S, Garny:2018byk, Givans:2020sez, Chen:2021rnb} have culminated in the EFT derivation of the one-loop power spectrum in Ref.~\cite{Ivanov:2023yla} and of the tree-level bispectrum
\cite{deBelsunceSullivanMcDonald2025}.

On large scales the \Lya forest flux contrast in redshift space can be expanded in powers of the linear density field $\delta_L$~\cite{Desjacques:2018pfv,Ivanov:2023yla,deBelsunce:2018xtd},
\begin{equation}
\label{eq:deltaF_PT}
\delta_F(\mathbf{k})=K_1(\mathbf{k})\,\delta_L(\mathbf{k})+\int_{\mathbf{q}_1,\mathbf{q}_2}(2\pi)^3\delta_D(\mathbf{k}-\mathbf{q}_1-\mathbf{q}_2)\,K_2(\mathbf{q}_1,\mathbf{q}_2)\,\delta_L(\mathbf{q}_1)\,\delta_L(\mathbf{q}_2)+\dots\,,
\end{equation}
with $\int_{\mathbf{q}}\equiv\int d^3q/(2\pi)^3$. The linear kernel is given by the Kaiser-like formula~\cite{McDonald:1999dt, McDonald:2001}
\begin{equation}
\label{eq:K1}
K_1(\mathbf{k})=b_1-b_\eta f\mu^2\,,
\end{equation}
where $\mu=k_\parallel/k$, $f$ is the linear growth rate, $b_1$ is the linear density bias and $b_\eta$ is the bias with respect to the line-of-sight gradient of the peculiar velocity.\footnote{We use the sign convention of the one-loop EFT Model \cite{Ivanov:2023yla}, which is also the one in our codes. With the opposite convention, $K_1=b_1+b_\eta f\mu^2$, $b_\eta$ and $b_{\delta\eta}$ change sign, while all other bias parameters are the same~\cite{deBelsunceSullivanMcDonald2025}.} The second-order kernel is
\begin{align}
\label{eq:K2}
K_2(\mathbf{q}_1,\mathbf{q}_2)={}&b_1\left[F_2(\mathbf{q}_1,\mathbf{q}_2)+\frac{f\mu_1\mu_2}{2}\left(\frac{q_2}{q_1}+\frac{q_1}{q_2}\right)\right]
-b_\eta\left[f\mu^2G_2(\mathbf{q}_1,\mathbf{q}_2)+\frac{f^2\mu_1\mu_2}{2}\left(\frac{q_2}{q_1}\mu_2^2+\frac{q_1}{q_2}\mu_1^2\right)\right]\nonumber\\
&+\frac{b_2}{2}+b_{\mathcal{G}_2}\left(\nu_{12}^2-1\right)-\frac{b_{\delta\eta}}{2}f\left(\mu_1^2+\mu_2^2\right)+b_{\eta^2}f^2\mu_1^2\mu_2^2\nonumber\\
&+b_{(KK)_\parallel}\left[\mu_1\mu_2\nu_{12}-\frac{\mu_1^2+\mu_2^2}{3}+\frac19\right]
+b_{\Pi^{[2]}_\parallel}\left[\mu_1\mu_2\nu_{12}+\frac57\mu^2\left(1-\nu_{12}^2\right)\right]\,,
\end{align}
where $\mu_i=q_{i\parallel}/q_i$, $\mu$ is the line-of-sight cosine of $\mathbf{q}_1+\mathbf{q}_2$, $\nu_{12}=\hat{\mathbf{q}}_1\cdot\hat{\mathbf{q}}_2$, and $F_2$ and $G_2$
are the standard perturbation theory kernels~\cite{Bernardeau:2001qr}. The terms proportional to $\mu_1\mu_2$ in the square brackets after $b_1$ and $b_\eta$ come from the line-of-sight displacement of the density and velocity-gradient fields by the redshift-space mapping. The other terms correspond to the second-order operators $\delta^2$, $\mathcal{G}_2$, $\delta\eta$, $\eta^2$, $(KK)_\parallel$ and $\Pi^{[2]}_\parallel$~\cite{Ivanov:2023yla}. For quasars (or galaxies), we can remap the bias parameters $b_\eta=-1$, $b_{\eta^2}=1$, $b_{\delta\eta}=-b_1$ and $b_{(KK)_\parallel}=b_{\Pi^{[2]}_\parallel}=0$ to obtain the standard redshift-space kernels.

\subsection{One-loop power spectrum}
\label{sec:p1loop}

In the EFT framework, the one-loop power spectrum is conceptually given by four components
\begin{equation}
\label{eq:Pmodel}
P(k,\mu)=P_{\rm tree}(k, \mu)+P_{\rm 1-loop}(k,\mu)+P_{\rm ct}(k,\mu)+P_{{\rm st}}(k,\mu)\,,
\end{equation}
which consist of the tree-level ($P_{\rm tree}$) Kaiser-like power spectrum scaling with the linear theory power spectrum ($P_{\rm lin}$), the one-loop correction ($P_{\rm 1-loop}$) which scales as $k^2P_{\rm lin}$ on large scales, the counterterms ($P_{\rm ct}$) which scale as $k^2P_{\rm lin}$ and stochastic components ($P_{\rm st}$) which consist of a constant and higher-order contributions. In the following, we will introduce the theory model for the auto-spectrum of the \Lya forest (denoted by $FF$) and the cross-spectrum between the \Lya forest and dark matter halos (as proxies for high-redshift galaxies or quasars; denoted by $FQ$).

For the auto-spectrum, the linear-bias term is given by the Kaiser-like formula~\cite{McDonald:1999dt, McDonald:2001}
\begin{equation}
\label{eq:PtreeFF}
P^{FF}_{\rm tree}(k,\mu) = K^2_1(\k) P_{\rm lin}(k)\,,
\end{equation}
where the linear kernel is given in Eq.~\eqref{eq:K1}.
The one-loop correction term is given by~\cite{Ivanov:2023yla}
\begin{equation}
\label{eq:P1loop}
P^{FF}_{\rm 1-loop}(k,\mu)=2\int_{\mathbf{q}}K_2^2(\mathbf{q},\mathbf{k}-\mathbf{q})P_{\rm lin}(|\mathbf{k}-\mathbf{q}|)P_{\rm lin}(q)+6K_1(\mathbf{k})P_{\rm lin}(k)\int_{\mathbf{q}}K_3(\mathbf{k},-\mathbf{q},\mathbf{q})P_{\rm lin}(q)\,,
\end{equation}
with the definition of the quadratic-order kernel, $K_2$, given in Eq.~\eqref{eq:K2}, and the third-order kernel $K_3$ in Ref.~\cite{Ivanov:2023yla} with five cubic bias parameters: $\{b_{\Gamma_3},b_{\Pi^{[3]}_\parallel},b_{\delta\Pi^{[2]}_\parallel},b_{(K\Pi^{[2]})_\parallel},b_{\eta\Pi^{[2]}_\parallel}\}$.\footnote{The cubic bias parameter $b_{\Gamma_3}$ is \emph{exactly} degenerate for the one-loop \Lya forest auto power spectrum stemming from a cubic kernel identity~\cite[see, Eq.~3.21 in Ref.][who set $b_{\Gamma_3}=0$]{Ivanov:2023yla}. 
This degeneracy is not lifted by the tree-level bispectrum so $b_{\Gamma_3}$ is prior-dominated in our analysis.
We keep this parameter explicitly to facilitate comparisons with analyses in which this degeneracy is broken, e.g., one-loop bispectrum or EFT-based analyses with simulation-based priors. } The counterterms are given by 
\begin{equation}
\label{eq:Pct}
P^{FF}_{\rm ct}(k,\mu)=-2\left(c_0+c_2\mu^2+c_4\mu^4\right)K_1(\mathbf{k})\,k^2P_{\rm lin}(k)\,,
\end{equation}
which absorb the dependence of the loop integrals on small, non-perturbative scales~\cite{McDonald:2009dh}, and the stochastic terms are
\begin{equation}
\label{eq:PstFF}
P^{FF}_{{\rm st}}(k,\mu)=P_{\rm shot}+\left(a_0+a_2\mu^2+a_4\mu^4\right)\frac{k^2}{\knl^2}\,.
\end{equation}
The constant $P_{\rm shot}$ and the $a_0$ and $a_2$ terms are the stochastic contributions of the one-loop model. The term $a_4\mu^4k^2$ comes from the non-locality in time of the dark matter dynamics and was not included in previous analyses~\cite{Ivanov:2023yla}. The non-linear scale is defined by $\knl$ which is calculated using $\Delta^2 = k^3P_{\rm lin}(k,z)/(2\pi^2)\sim 1$ for the snapshot at $z=2.5$: $k_{\rm NL} \approx 4.7 \hMpcinv$.

We use the same model for the cross-correlation with dark matter halos (as proxies for high-redshift galaxies and quasars) and replace one occurrence of the kernel with the one for quasars ($q$) at tree level and add additional cross-terms (denoted by $\times$)~\cite{Chudaykin:2025gsh},
\begin{align}
\label{eq:PFQ}
P_{FQ}(k,\mu)={}&K_1(\mathbf{k})K_1^q(\mathbf{k})P_{\rm lin}(k)+2\int_{\mathbf{q}}K_2(\mathbf{q},\mathbf{k}-\mathbf{q})K_2^q(\mathbf{q},\mathbf{k}-\mathbf{q})P_{\rm lin}(|\mathbf{k}-\mathbf{q}|)P_{\rm lin}(q)\nonumber\\
&+3P_{\rm lin}(k)\int_{\mathbf{q}}\left[K_1(\mathbf{k})K_3^q(\mathbf{k},-\mathbf{q},\mathbf{q})+K_1^q(\mathbf{k})K_3(\mathbf{k},-\mathbf{q},\mathbf{q})\right]P_{\rm lin}(q)\nonumber\\
&-\left[\left(c_0+c_2\mu^2+c_4\mu^4\right)K_1^q(\mathbf{k})+\left(c^q_0+c^q_2\mu^2+c^q_4\mu^4\right)K_1(\mathbf{k})\right]k^2P_{\rm lin}(k)\nonumber\\
&-b_4^{\times}f^4\mu^4k^4K_1(\mathbf{k})K_1^q(\mathbf{k})P_{\rm lin}(k)+P_{{\rm st}}^{FQ}(k,\mu)\,,
\end{align}
where $K_n^q$ are the redshift-space kernels of the quasars, with $K_1^q=b_1^q+f\mu^2$~\cite{Kaiser:1987qv}. They follow from the \Lya kernels for $b_\eta=-1$, $b_{\eta^2}=1$, $b_{\delta\eta}=-b_1^q$ and $b_{(KK)_\parallel}=b_{\Pi^{[2]}_\parallel}=0$, without the cubic \Lya operators. This leaves the quasar bias parameters $b_1^q$, $b_2^q$, $b_{\mathcal{G}_2}^q$ and $b_{\Gamma_3}^q$. The terms proportional to $b_4^q$ and $b_4^\times$ describe the fingers-of-God effect of the quasars at leading order~\cite{2020JCAP...05..042I,Chudaykin:2020hbf}.
 The stochastic contribution for the cross-spectrum is given by
\begin{equation}
\label{eq:PstFQ}
P_{{\rm st}}^{FQ}(k,\mu)=\left(a^\times_0+a^\times_2\mu^2+a^\times_4\mu^4\right)\frac{k^2}{\knl^2}\,.
\end{equation}

The quasar auto-spectrum is the standard one-loop model for galaxies in redshift space~\cite{Chudaykin:2020aoj},
\begin{equation}
\label{eq:PQQ}
P_{QQ}(k,\mu)=\left(K_1^q\right)^2P_{\rm lin}+P^q_{\rm 1-loop}-2\left(c^q_0+c^q_2\mu^2+c^q_4\mu^4\right)K_1^q\,k^2P_{\rm lin}-b_4^qf^4\mu^4k^4\left(K_1^q\right)^2P_{\rm lin}+P_{{\rm st},QQ}\,,
\end{equation}
and we use the same structure for the stochastic terms for the auto spectrum
\begin{equation}
\label{eq:PstQQ}
P_{{\rm st}}^{QQ}(k,\mu)=\frac{1}{\bar n_q}\left[P^q_{\rm shot}+\left(a^q_0+a^q_2\mu^2+a^q_4\mu^4\right)\frac{k^2}{\knl^2}\right]\,,
\end{equation}
where $1/\bar{n} = 1/(1.75 \times 10^{-4})[\Mpch]^{3}$ is the Poisson shot noise which we subtract from the measured spectrum, so that $P^q_{\rm shot}$ describes deviations from Poisson statistics. 

\subsubsection{One-loop IR resummation} 
In general, large-scale displacements damp the BAO wiggles which affects the \Lya forest in a similar way as galaxies. This results in an expression for the damping factor that is analogous to the one for galaxies~\cite{Ivanov:2023yla}. We split the linear power spectrum into a smooth and an oscillating part, $P_{\rm lin}=P_{\rm nw}+P_{\rm w}$, and split the one-loop power spectrum using time-sliced perturbation theory~\cite{Ivanov:2018gjr}
\begin{align}
\label{eq:irres}
P_{FF}={}&K_1^2\left[P_{\rm nw}+(1+\mathcal{S})e^{-\mathcal{S}}P_{\rm w}\right]+P_{\rm 1-loop}[P_{\rm nw}]+e^{-\mathcal{S}}\left(P_{\rm 1-loop}[P_{\rm lin}]-P_{\rm 1-loop}[P_{\rm nw}]\right)\nonumber\\
&+P_{\rm ct}\left[P_{\rm nw}+e^{-\mathcal{S}}P_{\rm w}\right]+P_{{\rm st},FF}\,,
\end{align}
where $X[P]$ denotes the term $X$ evaluated with the spectrum $P$. The damping function is
\begin{equation}
\label{eq:damp}
\mathcal{S}(k,\mu)=k^2\left[\Sigma^2\left(1+f\mu^2(2+f)\right)+\delta\Sigma^2f^2\mu^2\left(\mu^2-1\right)\right]\,,
\end{equation}
with
\begin{equation}
\Sigma^2=\frac{1}{6\pi^2}\int_0^{\Lambda_{\rm IR}}dq\,P_{\rm nw}(q)\left[1-j_0(qr_{\rm BAO})+2j_2(qr_{\rm BAO})\right]\,,\qquad
\delta\Sigma^2=\frac{1}{2\pi^2}\int_0^{\Lambda_{\rm IR}}dq\,P_{\rm nw}(q)\,j_2(qr_{\rm BAO})\,,
\end{equation}
where $j_0$ and $j_2$ are spherical Bessel functions, $\Lambda_{\rm IR}=0.2\hMpcinv$ and $r_{\rm BAO}=110\,\Mpch$. 
The same expressions hold for $P_{FQ}$ and $P_{QQ}$. 
We evaluate the model at the mean wavenumber of each $k$ bin and at the center of each $\mu$ bin.\footnote{We have verified that integrating over each bin instead does not affect our results.}

\subsection{Tree-level bispectrum}
\label{sec:btree}

The tree-level bispectrum is given by a deterministic (denoted by det), a stochastic (stoch) and their cross-term (mix) component
\be 
B_{\rm tree}=B_{\rm det}+B_{\rm mix}
+B_{\rm stoch}+B_{\rm h.d.}\,,
\ee 
where the first component is given by the standard expression
\begin{equation}
\label{eq:Btree}
B_{\rm det}(\mathbf{k}_1,\mathbf{k}_2,\mathbf{k}_3)=2K_1(\mathbf{k}_1)K_1(\mathbf{k}_2)K_2(\mathbf{k}_1,\mathbf{k}_2)P_{\rm lin}(k_1)P_{\rm lin}(k_2)+\text{2 cyclic permutations}\,,
\end{equation}
with $\mathbf{k}_1+\mathbf{k}_2+\mathbf{k}_3=0$. It depends on the eight bias parameters $\{b_1,b_\eta,b_2,b_{\mathcal{G}_2},b_{\delta\eta},b_{\eta^2},b_{(KK)_\parallel},b_{\Pi^{[2]}_\parallel}\}$~\cite{deBelsunceSullivanMcDonald2025}, which are the same parameters we sample in the one-loop power spectrum (Sec.~\ref{sec:p1loop};~\cite{Ivanov:2023yla}). The second part of the tree-level bispectrum is given by 
\be \label{eq:Bst}
B_{\rm mix}+B_{\rm stoch}=\frac{b_1(1+B_{\rm shot})+f\mu_1^2(1+C_{\rm shot})}{\bar n}K_1(\k_1)P_{\rm lin}(k_1)+\text{2 cyclic permutations}+\frac{1+A_{\rm shot}}{\bar n^2}\,,
\ee 
where $\bar n\sim 10~[\hMpcinv]^3$~\cite{deBelsunce:2025bqc, deBelsunce:2026tks}. 
The stochastic terms are split into multiple pieces. First, $C_{\rm shot}=P_{\rm shot}-2d_\eta (1+P_{\rm shot})$ which has two sources:
the RSD transport of the stochastic field $\epsilon$ which produces
$P_{\rm shot}$ as in the power spectrum 
plus an extra contribution
from a mixed operator 
$d_\eta \eta \epsilon\equiv \eta \epsilon_\eta$,
where $\epsilon_\eta$
is the line-of-sight (LOS) noise in the 
forest's rest frame. 
Since $d_\eta$ is unknown, 
$C_{\rm shot}$ should be treated as a
free parameter.
This term is also given as the last in Eq.~37 of Ref.~\cite{deBelsunceSullivanMcDonald2025}.
Further, we include a standard shot noise term $A_{\rm shot}$. 
The quadratic biases enter the power spectrum only through the loop corrections, while they enter the bispectrum already at tree level. In this work, by default, we only include the bispectrum of the \Lya forest itself, denoted by $FFF$. 
As an analysis variant we also investigate the quasar bispectrum ($QQQ$) which follows the same recipe as the above \Lya forest bispectrum except that we replace each occurrence of the \Lya kernel by the same one for quasars in redshift space~\cite{Chudaykin:2025aux}.

To account for the discreteness of the measured bispectrum which averages over Fourier modes in the box  we follow Ref.~\cite{Ivanov:2021kcd} and apply the integral approximation and correct for the discreteness of the Fourier modes. Its expectation value in a bin with sides in the shells $k_1$, $k_2$ and $k_3$ is the average of the model over all closed triangles of grid wavevectors in the bin given by
\begin{equation}
\label{eq:Bmodbin}
B^{\rm grid}_\ell(k_1,k_2,k_3)=\frac{2\ell+1}{N_T}\sum_{\mathbf{q}_1\in k_1}\sum_{\mathbf{q}_2\in k_2}\sum_{\mathbf{q}_3\in k_3}\delta^{\rm K}(\mathbf{q}_1+\mathbf{q}_2+\mathbf{q}_3)\,B(\mathbf{q}_1,\mathbf{q}_2,\mathbf{q}_3)\,\mathcal{L}_\ell(\mu_3)\,,
\end{equation}
where $N_T$ is the number of closed triangles in the bin and $\mu_3$ is the cosine of the angle between the longest side and the line of sight. 
We only evaluate this expression once. First, we evaluate the continuum limit, i.e.,~we assume that the Fourier modes are infinitely dense, so that the sum over the modes becomes an integral:
\begin{equation}
\label{eq:Bcont}
B^{\rm int}_\ell(k_1,k_2,k_3)=\frac{2\ell+1}{
V_T}\prod_{i=1}^{3}\left[\int_{k_i}\!dq_i\,q_i\right]\int_0^{2\pi}\frac{d\phi}{2\pi}\int_{-1}^{1}\frac{d\mu}{2}\,\mathcal{I}(q_1,q_2,q_3)\,B(q_1,q_2,q_3,\mu_1,\mu_2,\mu_3)\,\mathcal{L}_\ell(\mu)\,,
\end{equation}
where $\mathcal{I}$ is one if $q_1$, $q_2$ and $q_3$ form a closed triangle and zero otherwise, $\mu$ is again the observed cosine of the angle between the longest side and the line of sight, $\phi$ is the angle of the triangle around its longest side, and $V_T$ is the same integral without $B\,\mathcal{L}_\ell$ and the volume factor.
Our treatment of binning at this stage is similar to Refs.~\cite{Chudaykin:2025aux, Chudaykin:2025vdh, Ivanov:2026dvl, Bakx:2025pop}.
Note, however, that the continuum limit is not exact in a finite box. 
Second, we therefore correct for the difference between Eqs.~\eqref{eq:Bmodbin} and \eqref{eq:Bcont}, which we compute once at the fiducial cosmology. Ref.~\cite{Ivanov:2021kcd} applies this correction as the ratio of the two total bispectra (the so-called discreteness weights). We add it to each term of the model separately, so that the model equals the sum over the grid triangles at the fiducial cosmology for any values of the parameters.
Our final theory expression is given by 
\begin{equation}
\label{eq:Bth}
B^{\rm th}_\ell(k_1,k_2,k_3)=\sum_j c_j\left[B^{\rm int}_{\ell,j}(k_1,k_2,k_3)+\left(B^{\rm grid}_{\ell,j}-B^{\rm int}_{\ell,j}\right)_{\rm fid}(k_1,k_2,k_3)\right]\,,
\end{equation}
where $c_j$ runs over the products of parameters that multiply the terms of the model, $B^{\rm int}_\ell$ is computed from Eq.~\eqref{eq:Bcont} using our bispectrum model and $B_\ell^{\rm grid}$  is the same term summed over the grid triangles, and ``fid'' denotes the fiducial cosmology. The correction accounts for the difference between the continuum limit and the sum over the finite set of grid modes, Eq.~\eqref{eq:Bmodbin}.

\subsubsection{IR resummation}
We include the damping of the BAO wiggles at leading order by replacing the linear power spectrum on each side by
\begin{equation}
\label{eq:Bir}
P_{\rm lin}(k_i)\to P_{\rm nw}(k_i)+e^{-\mathcal{S}(k_i,\mu_i)}P_{\rm w}(k_i)\,,
\end{equation}
where $P_{\rm nw}$ and $P_{\rm w}$ are the smooth and oscillating parts of the linear power spectrum, and $\mathcal{S}$ is the damping function of Eq.~\eqref{eq:damp}, with the same $\Sigma^2$ and $\delta\Sigma^2$ as in the power spectrum. In the one-loop power spectrum the tree-level wiggles carry an extra factor $(1+\mathcal{S})$, which avoids counting the damping already contained in the one-loop term twice. Since the bispectrum has no loop term here, this factor is absent.

\subsubsection{Higher-derivative terms} 
On small scales the linear response of the flux receives corrections suppressed by $k^2$, from higher-derivative terms in the bias expansion such as $\nabla^2\delta$ and $\partial_\parallel^2\delta$. To this order they change the linear kernel to
\begin{equation}
\label{eq:K1ct}
K_1(\mathbf{k})\to K_1(\mathbf{k})-C(\mu)\,k^2\,,\qquad C(\mu)=c^B_0+c^B_2\mu^2+c^B_4\mu^4\,.
\end{equation}
In the power spectrum the same change gives the counterterms of Eq.~\eqref{eq:Pct}. In the bispectrum, $K_1$ appears once on each side of the triangle, and each side gets its own correction, evaluated at its $k_i$ and $\mu_i$. To first order in the coefficients,
\begin{equation}
\label{eq:Bct}
B_{\rm h.d.}(\mathbf{k}_1,\mathbf{k}_2,\mathbf{k}_3)=-2\left[C(\mu_1)k_1^2K_1(\mathbf{k}_2)+K_1(\mathbf{k}_1)C(\mu_2)k_2^2\right]K_2(\mathbf{k}_1,\mathbf{k}_2)P_{\rm lin}(k_1)P_{\rm lin}(k_2)+\text{2 cyclic permutations}\,,
\end{equation}
again with the damped spectrum of Eq.~\eqref{eq:Bir} on each side. Formally these terms are of the same order as the one-loop corrections. As for galaxies~\cite{Chudaykin:2025aux}, we include them at tree level, where they also absorb part of the one-loop bispectrum that we do not Model \cite{Ivanov:2021kcd}. Our model is thus (partially) a one-loop marginalized bispectrum which we decide to include in the model to be more conservative. For the monopole of the galaxy bispectrum the $\mu^2$ term is sufficient, while higher multipoles also need the $\mu^4$ term~\cite{Ivanov:2021kcd}. Since we fit both $B_0$ and $B_2$, we keep all three powers of $\mu$, as in the power spectrum.

Eq.~\eqref{eq:Btree} is quadratic in $b_1$ and $b_\eta$ (through $K_1K_1$) and linear in the eight parameters of $K_2$, and Eq.~\eqref{eq:Bct} is linear in the counterterms, in $b_1$ and $b_\eta$, and in the parameters of $K_2$. For a given linear power spectrum and growth rate, we therefore compute the bin average of Eq.~\eqref{eq:Bth} once for each product of parameters. Evaluating the model at a new set of parameters is then a matrix--vector product.

\subsubsection{Bispectrum covariance}
\label{sec:Bcov}

We use the Gaussian covariance of the estimator in Eq.~\eqref{eq:Bestimator}. For a triangle bin $t$,
\begin{equation}
\label{eq:Bcov}
{\rm Cov}\left[\hat B_\ell,\hat B_{\ell'}\right](t)=\frac{(2\ell+1)(2\ell'+1)\,V}{N_T}\left\langle P(\mathbf{q}_1)P(\mathbf{q}_2)P(\mathbf{q}_3)\,\mathcal{L}_\ell(\mu_1)\sum_{\pi}\mathcal{L}_{\ell'}(\mu_{\pi(1)})\right\rangle_t\,,
\end{equation}
where $\langle\cdot\rangle_t$ is the average over the closed triangles of the bin. The sum runs over the permutations $\pi$ of the three sides that map the bin onto itself: only the identity for scalene bins, two permutations for isosceles bins and six for equilateral bins. For $\ell=\ell'=0$ and an isotropic spectrum, Eq.~\eqref{eq:Bcov} reduces to the familiar $s_t\,V\,P(k_1)P(k_2)P(k_3)/N_T$ with $s_t=1$, $2$ and $6$~\cite{Scoccimarro:2000sn,Sefusatti:2006pa}. 
The covariance couples the multipoles within each bin, but different bins are uncorrelated. We take $N_T$ from the estimator, and we evaluate the average over the same triangles as the model.

For the spectrum on each side we use the tree-level model, $P(\mathbf{q})=K_1^2(\mathbf{q})P_{\rm lin}(q)$, with $b_1$ and $b_\eta$ fixed to the best fit of the one-loop power spectrum, as for galaxies~\cite{Ivanov:2021kcd}.
We neglect the non-Gaussian contributions to the covariance and the cross-covariance between the power spectrum and the bispectrum, which is small for galaxies~\cite{Ivanov:2021kcd} on large scales. When we fit the mean of $N_{\rm sim}$ boxes, we divide the covariance again by six. We leave to future work to create simulation-based covariance matrices using field-level simulations and investigating the cross-covariance between power spectrum and bispectrum~\cite{deBelsunce:2026tks}.

\subsection{Alcock--Paczy\'nski effect}\label{sec:EFT_AP}
We measure the power spectrum and the bispectrum in the comoving coordinates of the simulation, which are set by its input cosmology. For observational data, however, one needs to assume a fiducial cosmology to convert redshifts and angles into distances. To account for distortions stemming from a mismatch between the fiducial and ``true'' cosmological 
parameters the Alcock--Paczy\'nski (AP) effect was introduced~\citep{AlcockPaczynski1979}. Therefore, we introduce the scaling parameters
\begin{equation}
\label{eq:alpha_AP}
\qpar\equiv\frac{H^{\rm fid}(z)}{H(z)}\,,\qquad
\qperp\equiv\frac{D_A(z)}{D^{\rm fid}_A(z)}\,,
\end{equation}
where $H(z)$ is the Hubble parameter and $D_A(z)$ the angular diameter distance at the redshift of the \abacus mock evaluated at the sampled cosmology and the superscript ``fid'' denotes the same parameters evaluated at the fiducial cosmology, for which $\qpar=\qperp=1$ at the true cosmology. 
Both parameters are derived from the cosmological parameters at each step of the chain, and we report them in App.~\ref{sec:tab_meandv_freecosmo} for mean data vector fits.  An observed wavevector is rescaled to 
\be 
k\rightarrow k' = k\left[\frac{\mu^2}{\qpar^2}+\frac{1-\mu^2}{\qperp^2}\right]^{1/2}\,, \qquad \mu \rightarrow \mu' = \frac{\mu}{\qpar}\left[\frac{\mu^2}{\qpar^2}+\frac{1-\mu^2}{\qperp^2}\right]^{-1/2}\,.
\ee
This corresponds to fitting the power spectrum predictions as a function of $k'$ and $\mu'$, \textit{i.e.}, 
\begin{align}\label{eq:Pmodtogrid}
    P(k,\mu)&= \frac{1}{\qperp^2\qpar} P(k',\mu')\,,
\end{align}
and the tree-level bispectrum predictions as a function of the re-mapped sides of the triangle in Eq.~\eqref{eq:Bcont},
\begin{align}\label{eq:Bmodtogrid}
    B(k_1,k_2,k_3,\mu_1,\mu_2,\mu_3)&= \frac{1}{(\qperp^2\qpar)^2}\, B(k'_1,k'_2,k'_3,\mu'_1,\mu'_2,\mu'_3)\,,
\end{align}
where each side $(k_i,\mu_i)$ is re-mapped. For full-shape fits when varying the cosmological parameters we apply the dilation to the entire theory model and normalize it.

\section{Likelihood and priors}
\label{sec:like}

We fit the power spectrum and the bispectrum of the mean of the boxes with a Gaussian likelihood,
\begin{equation}
\label{eq:like}
-2\ln\mathcal{L}=\left(\mathbf{d}_P-\mathbf{t}_P\right)^{\rm T}\mathbf{C}_P^{-1}\left(\mathbf{d}_P-\mathbf{t}_P\right)+\left(\mathbf{d}_B-\mathbf{t}_B\right)^{\rm T}\mathbf{C}_B^{-1}\left(\mathbf{d}_B-\mathbf{t}_B\right)\,,
\end{equation}
where $\mathbf{d}_P$ is the measured power spectrum in bins of $k$ and $\mu$, $\mathbf{d}_B$ contains the bispectrum multipoles $B_0$ and $B_2$, and $\mathbf{t}_P$ and $\mathbf{t}_B$ are the models of Sec.~\ref{sec:theory}. We neglect the cross-covariance between the two statistics, so that the likelihood is the product of the likelihoods of the power spectrum and of the bispectrum. For the auto power spectrum we use the diagonal Gaussian covariance $C_{ii}=2P_i^2/N_i$, where $N_i$ is the number of modes in bin $i$.  For the cross-spectrum covariance ($FQ$) we use the form $C_{ii}^{\rm FQ} = N_{i}^{-1}\left[(P_{i}^{FQ})^2+P_i^{FF}P_i^{QQ}\right]$ where the quasar-quasar power spectrum is denoted by $P^{\rm QQ}$ which includes the shot noise $1/\bar{n} = 1/(1.75 \times 10^{-4})[\Mpch]^{3}$~\cite{Hadzhiyska:2025cvk}.
Note that we ignore the cross-covariance between the flux and quasars in the power spectrum, i.e.,~between $FF$, $FQ$ and $QQ$. In principle, the inclusion of the cross-covariance would lead to cosmic variance cancellation which can additionally tighten constraints~\cite{Chudaykin:2025gsh}, yet for the present work we choose to be more conservative and use analytic covariance matrices whose accuracy for the cross-covariances have not been fully validated.
For the bispectrum we use the expression for the covariance given in Eq.~\eqref{eq:Bcov} with the tree-level power spectrum.
We use the power spectrum up to $k_{\max}=0.6\,(0.3)\,\hMpcinv$ for the auto- (cross-) spectrum, $k_{\max}^{QQ}=0.2\,\hMpcinv$ for the quasar auto-spectrum (if included) and the bispectrum up to $k_{\max}^B=0.1\hMpcinv$ (our baseline analysis choice), and, as a conservative choice, up to $k_{\max}^B=0.06\hMpcinv$.

For the present analysis, we vary the cosmological parameters $ \{\omega_{cdm},  \lnAs, h \}$ with $H_0,\, \sigma_8,\, \Omega_m$ as derived parameters and fix $n_s$ and $\omega_b$ to the fiducial values of the \textsc{AbacusSummit} simulations. We list all parameters and their priors in Tab.~\ref{tab:params_priors}. For the parameters of the one-loop power spectrum we use the same priors as in previous work~\cite{deBelsunce:2024rvv}. We explore the cosmological and EFT parameter space using the Markov Chain Monte Carlo (MCMC) sampler \texttt{Cobaya}~\footnote{Publicly available at \url{https://github.com/CobayaSampler/cobaya}.}~\citep{Torrado:2020dgo}. To speed up the generation of the linear matter power spectrum, we use the \texttt{jaxmapse}~\footnote{Publicly available at \url{https://github.com/CosmologicalEmulators/jaxmapse}.} emulator, which uses the same backend used in~\cite{Bonici:2023xjk, Bonici:2025ltp}
and implement the FFTLog technique to evaluate the EFT kernels~\cite{Chudaykin:2020aoj, Simonovic:2017mhp} on CUDA kernels. Each fit to an \abacus simulation takes one GPU-node hour (using one NVIDIA A100 GPU on the Perlmutter computer at NERSC) and is considered converged once the Gelman-Rubin diagnostic reaches $\hat R -1 < 0.02$ for all parameters among eight chains~\citep{1992StaSc...7..457G}. 

To mitigate the effect of projection (or prior volume) effects, we directly sample the quantity constrained by the data: $b_{\mathcal{O}}\sig^{\mathcal{O}}$ where for the linear (quadratic) bias parameters the order is $\mathcal{O}=1$ ($\mathcal{O}=2$). Additionally, for the RSD parameter we sample $b_{\eta}\sig,\, b_{\eta 2}\sig^2, b_{\delta \eta}\sig^2$ (see, e.g.,~\cite{Chudaykin:2025aux}) with the priors given in Tab.~\ref{tab:params_priors}.
\begin{table*}
    \centering
    \footnotesize
    \setlength{\tabcolsep}{2.5pt}
    \renewcommand{\arraystretch}{1.08}
    \begin{tabular}{@{}ll@{\hspace{1.0em}}ll@{\hspace{1.0em}}ll@{\hspace{1.0em}}ll@{}}
        \hline\hline
        \multicolumn{2}{c}{{Cosmological}} &
        \multicolumn{2}{c}{{Sampled EFT}} &
        \multicolumn{2}{c}{{Analytically marginalized}} &
        \multicolumn{2}{c}{{Analytically marginalized}} \\
        \multicolumn{2}{c}{{parameters}} &
        \multicolumn{2}{c}{{parameters}} &
        \multicolumn{2}{c}{$P_{FF}$} &
        \multicolumn{2}{c}{$P_{FQ}$, $P_{QQ}$, and $B_{FFF}$} \\
        \hline
        Parameter & Prior &
        Parameter & Prior &
        Parameter & Prior &
        Parameter & Prior \\
        \hline
        $H_0\,[{\rm km\,s^{-1}\,Mpc^{-1}}]$ &
        $\mathcal{U}(60,75)$ &
        $b_1\sigma_8$ &
        $\mathcal{U}(-2,2)$ &
        $A b_{\Gamma_3}$ &
        $\mathcal{N}(0,10^2)$ &
        $A b_{\Gamma_3}^q$ &
        $\mathcal{N}(0,5^2)$ \\

        $\omega_{\rm cdm}$ &
        $\mathcal{U}(0.08,0.16)$ &
        $b_\eta\sigma_8$ &
        $\mathcal{U}(-2,2)$ &
        $A b_{(K\Pi^{[2]})_\parallel}$ &
        $\mathcal{N}(0,2^2)$ &
        $A c^q_{s,0},\,A c^q_{s,2},\,A c^q_{s,4}/[h^{-1}{\rm Mpc}]^2$ &
        $\mathcal{N}(0,10^2)$ 
        \\

        $\ln(10^{10}A_s)$ &
        $\mathcal{U}(2.03,3.93)$ &
        $b_2\sigma_8^2$ &
        $\mathcal{N}(0,2^2)$ &
        $A b_{\delta\Pi^{[2]}_\parallel}$ &
        $\mathcal{N}(0,2^2)$ &
        $A b_{4,\times}/[h^{-1}{\rm Mpc}]^4$ &
        $\mathcal{N}(0,100^2)$ 
        \\

        & &
        $b_{\mathcal{G}_2}\sigma_8^2$ &
        $\mathcal{N}(0,2^2)$ &
        $A b_{\eta\Pi^{[2]}_\parallel}$ &
        $\mathcal{N}(0,2^2)$ &
        $Aa_{0}^{\times}/P_\epsilon\,, Aa_{2}^{\times}/P_\epsilon\,, Aa_{4}^{\times}/P_\epsilon$ &
        $\mathcal{N}(0,1^2)$ 
        \\

        & &
        $b_{(KK)_\parallel}\sigma_8^2$ &
        $\mathcal{N}(0,2^2)$ &
        $A b_{\Pi^{[3]}_\parallel}$ &
        $\mathcal{N}(0,2^2)$ &
        $A b_4^q/[h^{-1}{\rm Mpc}]^4$ &
        $\mathcal{N}(0,100^2)$ 
        \\

        & &
        $b_{\Pi^{[2]}_\parallel}\sigma_8^2$ &
        $\mathcal{N}(0,2^2)$ &
        $A c_{0,2,4}/[h^{-1}{\rm Mpc}]^2$ &
        $\mathcal{N}(0,1^2)$ &
        $\bar n_q P_{\rm shot}^q$ &
        $\mathcal{N}(0,1^2)$ 
        \\

        & &
        $b_{\delta\eta}\sigma_8^2$ &
        $\mathcal{N}(0,2^2)$ &
        $P_{\rm shot}/P_\epsilon$ &
        $\mathcal{N}(0,5^2)$ &
        $\bar n_q a_0^q,\,\bar n_q a_2^q,\,\bar n_q a_4^q$ &
        $\mathcal{N}(0,1^2)$ 
        \\

        & &
        $b_{\eta^2}\sigma_8^2$ &
        $\mathcal{N}(0,2^2)$ &
        $a_{0,2}/P_\epsilon$ &
        $\mathcal{N}(0,5^2)$ &
        $A c^B_{0},\,A c^B_{2}/[h^{-1}{\rm Mpc}]^2$ &
        $\mathcal{N}(0,1^2)$ 
        \\

        & &
        $b_1^q\sigma_8$ &
        $\mathcal{U}(0,10)$ &
        $a_4/P_\epsilon$ &
        $\mathcal{N}(0,10^2)$ &
        $A c^B_{4}/[h^{-1}{\rm Mpc}]^2$ &
        $\mathcal{N}(0,10^2)$ 
        \\

        & &
        $b_2^q\sigma_8^2$ &
        $\mathcal{U}(-10,10)$ &
        & &
        $B_{\rm shot},\,C_{\rm shot},\,A_{\rm shot}$ &
        $\mathcal{N}(0,1^2)$ 
        \\

        & &
        $b_{\mathcal{G}_2}^q\sigma_8^2$ &
        $\mathcal{U}(-10,10)$ &
        & &
        \\
        \hline\hline
    \end{tabular}
    \caption{Cosmological and EFT parameters and priors adopted for the joint analysis of the \Lya forest auto-power spectrum, $P_{FF}$, its cross-power spectrum with quasars, $P_{FQ}$, the quasar auto-power spectrum, $P_{QQ}$, and the tree-level bispectrum, $B_{FFF}$. 
    The last two panels list the parameters that enter the likelihood linearly and are marginalized analytically, for $P_{FF}$ and for $P_{FQ}$, $P_{QQ}$ and $B_{FFF}$. The stochastic amplitudes of the \Lya forest power spectra are in units of $P_\epsilon=2\pi^2/\knl^3$, those of the quasar auto-spectrum in units of the quasar shot noise, $1/\bar n_q\approx5714\,(\Mpch)^3$, and those of the bispectrum are defined in Eq.~\eqref{eq:Bst}, with $\bar n=10\,[h\,{\rm Mpc}^{-1}]^3$. We use $A\equiv\sigma_8^2(z)/\sigma_{8,{\rm ref}}^2(z)$, where $\sigma_{8,{\rm ref}}(z)$ is the fluctuation amplitude at the fiducial cosmology of the simulation, to reduce projection effects~\cite{Tsedrik:2025hmj}. We fix $\omega_b=0.02237$ and $n_s=0.9649$ to the fiducial \abacus cosmology.}
    \label{tab:params_priors}
\end{table*}

\section{Results}\label{sec:results}
In this section we apply the EFT theory models, introduced in Sec.~\ref{sec:theory}, to the \Lya forest mocks, introduced in Sec.~\ref{sec:mocks}, and evaluate their performance using as key metric the unbiased recovery of the cosmological parameters and constraints on the EFT nuisance parameters. We directly sample the cosmological parameters and perform full-shape fits in Sec.~\ref{sec:res_free_cosmo} and quantify the improvement in cosmological constraints obtained by adding the tree-level \Lya forest bispectrum ($FFF$) to the one-loop power spectrum ($FF$ and $FQ$). 
We show the mean bispectrum data vector in Sec.~\ref{sec:res_bk_dv} for the \Lya forest and discuss the theory modeling for the tree-level bispectrum. 
Finally, we add the quasar auto-power spectrum and bispectrum to the baseline fits from Sec.~\ref{sec:res_free_cosmo} in Sec.~\ref{sec:Pqq_Bqqq}. 
Our baseline analysis choice follows the tracers used for the DESI analyses~\cite{DESI:2026lnd} and uses the \Lya forest auto and cross-spectrum between the \Lya forest and quasars ($FF$+$FQ$). As scale cuts we impose $k_{\rm max}^{FF\,(FQ)} = 0.6\, (0.3)\hMpcinv$ and, if included, the \Lya forest bispectrum ($FFF$) with $k_{\rm max}^{FFF}=0.1\hMpcinv$. 
As an analysis variation we also investigate the impact of adding the quasar auto power spectrum ($QQ$) with $k_{\rm max}^{QQ}=0.2 \hMpcinv$ and bispectrum ($QQQ$) with $k_{\rm max}^{QQQ}=0.08 \hMpcinv$~\cite{Chudaykin:2025aux}. 
We discuss the chosen scale cuts of $FF$, $FQ$ and $FFF$ from fits to the mean data vector of the 12 mocks in App.~\ref{sec:cosmo_scale_cuts}.

\subsection{Free-cosmology runs}\label{sec:res_free_cosmo}
\begin{figure}
    \centering
    \includegraphics[width=0.49\linewidth]{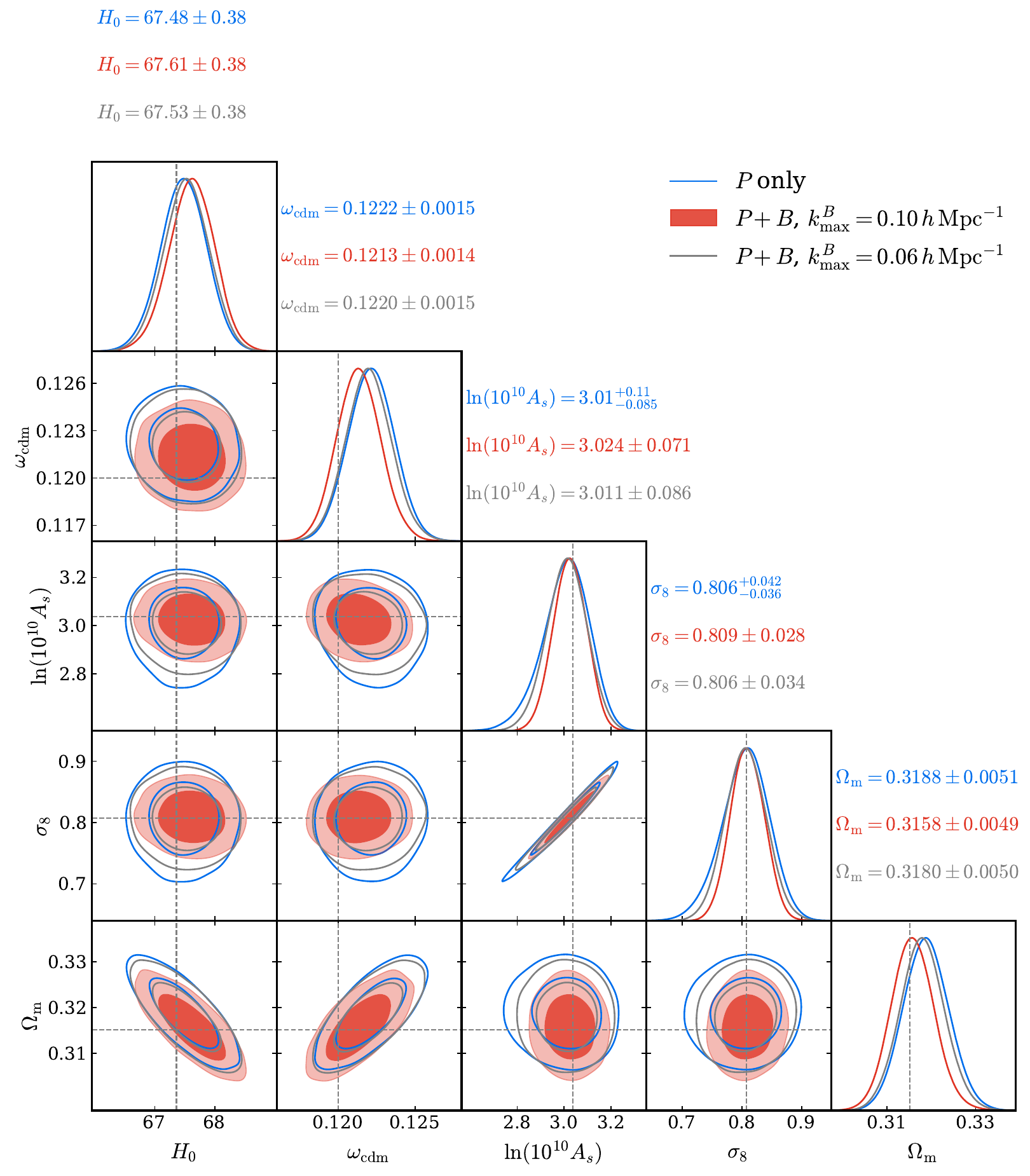}\hfill 
    \includegraphics[width=0.49\linewidth]{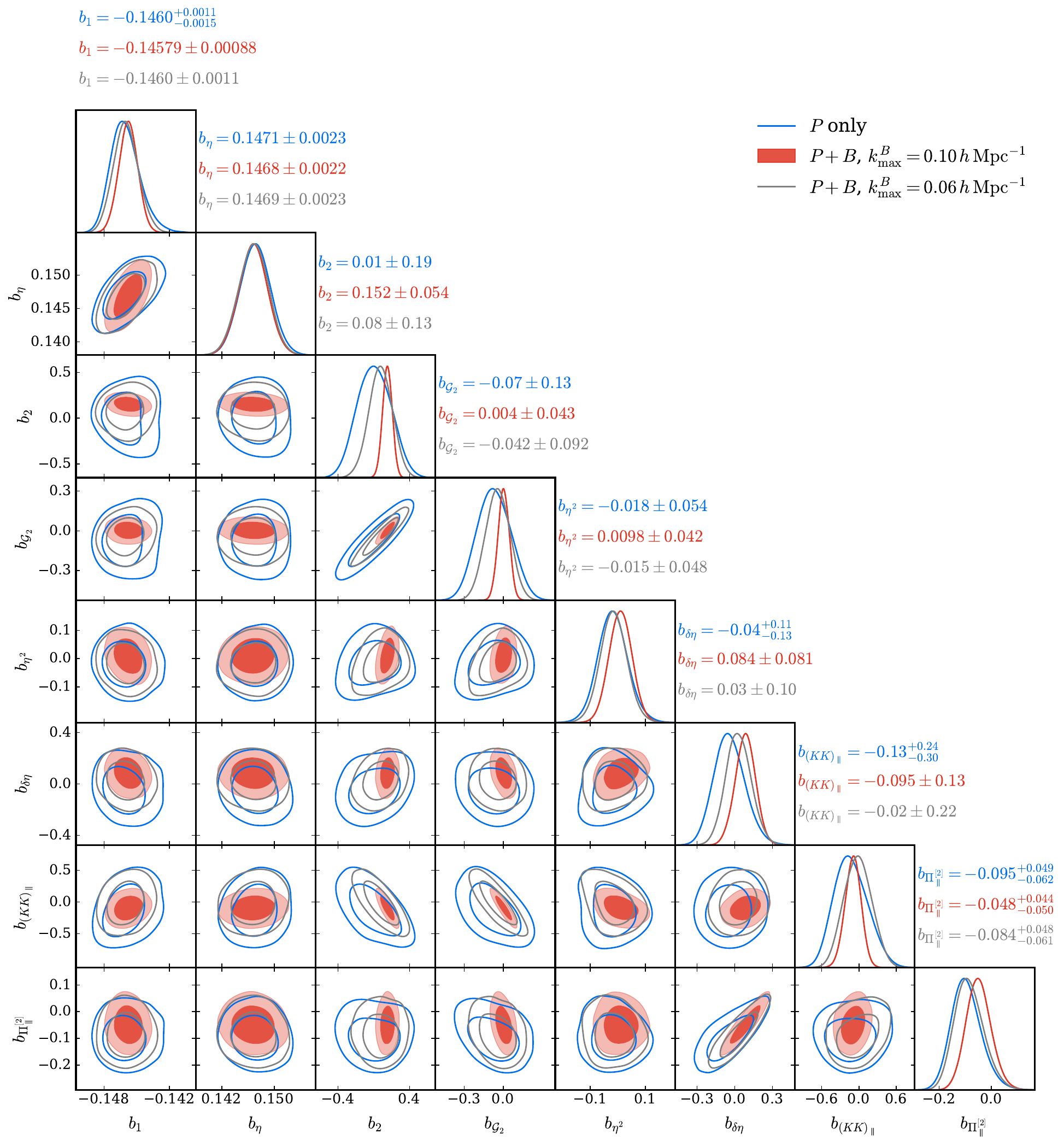}
    \vspace{-0.1in}
    \caption{\textbf{Bispectrum improves constraints on cosmological parameters.} We compare the marginalized 1D (68\% C.L.) and 2D (68\% and 95\% C.L.) constraints on the cosmological (left) and nuisance (right panel) parameters obtained from fits to the \abacus simulation (Model 1, phase 1, LOSz) using the power spectrum ($\kmax=0.6 \hMpcinv$) alone (blue) and the power spectrum jointly with the tree-level bispectrum using two scale cuts of $\kmax=0.06 \hMpcinv$ (gray; conservative) and $\kmax=0.10 \hMpcinv$ (red; baseline). We recover unbiased cosmological parameters from both the $P$ and $P+B$ fits. Crucially, the bispectrum improves the constraints on $\sig$ by 28\% (13\%) and on $\lnAs$ by 27\% (12\%) using the baseline (conservative) scale cut of $k_{\rm max}^B=0.1\, (0.06)\hMpcinv$. We find no improvement for $H_0$ and $\omega_{\rm cdm}$. Including the tree-level bispectrum also improves the constraints on the linear bias parameter, $b_1$, and the quadratic bias parameters. We tabulate the results for the cosmological parameters in Tab.~\ref{tab:constraints_H0_omega_cdm_ln1e10As_sigma8}. The dashed gray lines indicate the fiducial input cosmology of the \abacus mock.
    }
    \label{fig:freecosmo_PB}
\end{figure}

In this section, we sample the cosmological parameters in addition to the bias parameters for a single simulation realization (fiducial \abacus cosmology, phase 1, Model 1). The power spectrum uses $k_{\max}^{FF}=0.6 \hMpcinv$ for the auto-correlation and $k_{\max}^{FQ}=0.3 \hMpcinv$ for the cross-correlation with quasars. For the bispectrum ($FFF$) we compare two scale cuts with $\kmax=0.06$ and $0.10 \hMpcinv$ illustrated as gray lines and red contours, respectively. 
We show the marginalized 1D and 2D posteriors of the recovered cosmological parameters (left panel) and leading order bias parameters (right panel) in Fig.~\ref{fig:freecosmo_PB}. We find excellent recovery of the cosmological parameters at the $\approx 1\sigma$-level. The strongest improvements are obtained for  $\sig$ by  28\% and $\lnAs$ by 27\%. For the bias parameters, adding the bispectrum improves the constraints on $b_2$ by 72\%, on $b_{\mathcal{G}_2}$ by 67\%, on $b_{(KK)_\parallel}$ by 52\% and on the linear bias parameter, $b_1$, by 32\%. The bispectrum breaks the degeneracy between the mean flux and overall power spectrum normalization (and some higher order bias parameters) which does not exist for $b_\eta$. The latter does thus not improve when including the bispectrum, as expected. We do not show the cubic parameters in the triangle plot as these are only loosely constrained by the employed models and adding the tree-level bispectrum does not improve their constraints, as expected. In Tab.~\ref{tab:constraints_H0_omega_cdm_ln1e10As_sigma8} we summarize the posterior means and 1D marginalized error bars (68\% C.L.) for the fits for the cosmological ($H_0,\, \omega_{\rm cdm},\, \lnAs$) and derived parameters ($\sig,\, \Omega_{\rm m}$).
We extend this analysis to investigate the robustness to different noise realizations (i.e.,~phases) in App.~\ref{sec:app_phases} for all the cosmological and EFT parameters.

\begin{table*}
    \centering
    \setlength{\tabcolsep}{16pt}
    \renewcommand{\arraystretch}{1.2}
    \begin{tabular}{@{}lccc@{}}
    \hline
    Parameter & \multicolumn{1}{c}{$P$ only} & \multicolumn{2}{c}{$P+B$} \\
    \cline{3-4}
              &                              & $k_{\max}^{B}=0.10\,h\,{\rm Mpc}^{-1}$
                                             & $k_{\max}^{B}=0.06\,h\,{\rm Mpc}^{-1}$ \\
    \hline
        ${H_0}$ &
        $67.48\pm0.38$ &
        $67.61\pm0.38$ &
        $67.53\pm0.38$ \\

        ${\omega_{\rm cdm}}$ &
        $0.1222\pm0.0015$ &
        $0.1213\pm0.0014$ &
        $0.1220\pm0.0015$ \\

        $\ln(10^{10}A_s)$ &
        $3.01^{+0.11}_{-0.085}$ &
        $3.024\pm0.071$ &
        $3.011\pm0.086$ \\

        $\sigma_8$ &
        $0.806^{+0.042}_{-0.036}$ &
        $0.809\pm0.028$ &
        $0.806\pm0.034$ \\

        $\Omega_{\rm m}$ &
        $0.3188\pm0.0051$ &
        $0.3158\pm0.0049$ &
        $0.3180\pm0.0050$ \\
        \hline\hline
    \end{tabular}
    \caption{\textbf{Results on free-cosmology fits on single mock:} 
    1D marginalized constraints ($68\%$ C.L.) for the parameters shown in Fig.~\ref{fig:freecosmo_PB}. The top three rows show the sampled parameters with $\sigma_8$ and $\Omega_{\rm m}$ being derived parameters. We compare the addition of the \Lya forest bispectrum using two different scale cuts: (i) baseline ($k_{\max}^{B}=0.1\,h\,{\rm Mpc}^{-1}$) and (ii) conservative $k_{\max}^{B}=0.06\,h\,{\rm Mpc}^{-1}$.  We find excellent agreement between all three fits with 27 (12)\% improvement on $\lnAs$ and 28 (13)\% improvement on $\sig$ for the addition of the bispectrum using the baseline (conservative) scale cut compared to the power spectrum alone. 
    }
    \label{tab:constraints_H0_omega_cdm_ln1e10As_sigma8}
\end{table*}

\begin{figure}
    \centering
    \includegraphics[width=.8\linewidth]{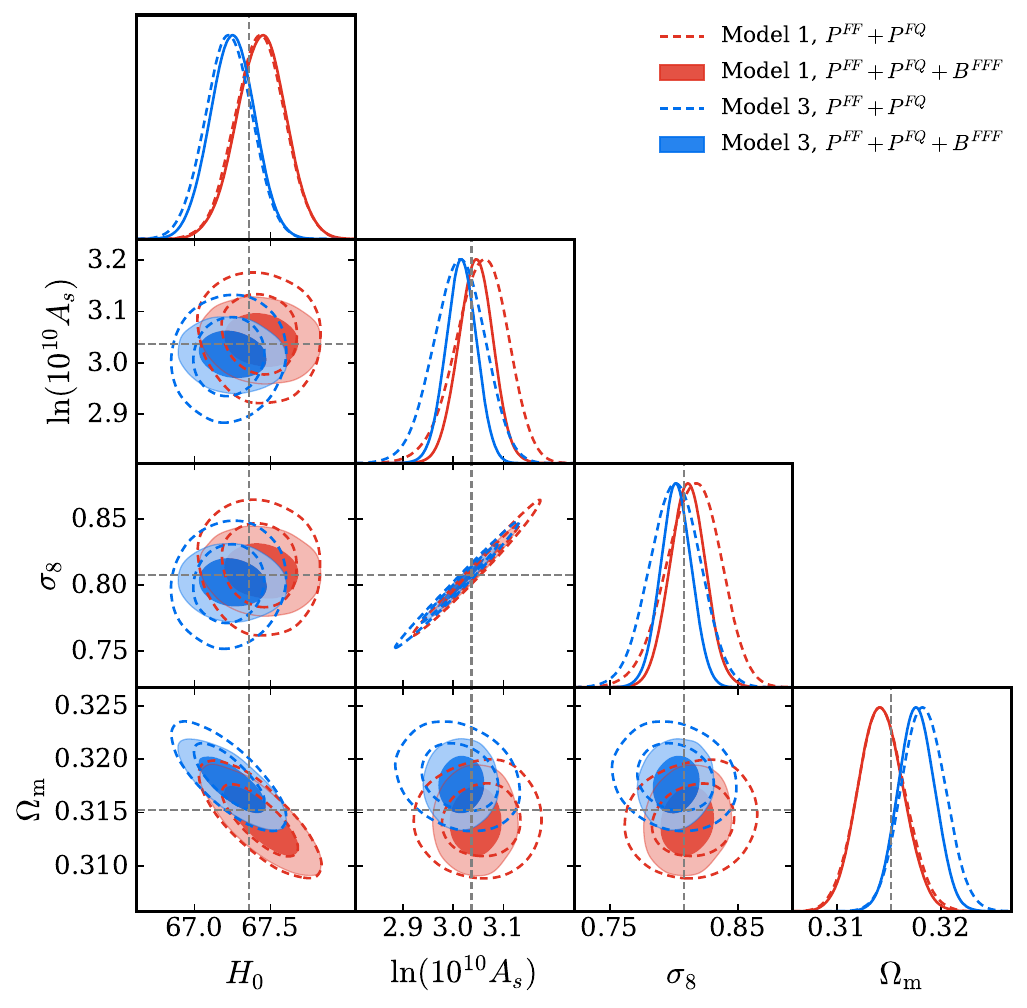}
    \vspace{-0.2in}
    \caption{\textbf{Comparison of mean data vector fits:} Fits to the mean data vector over all available 12 realizations (six phases and two lines of sight) and dividing the covariance matrix by a factor of 12 for models 1 (red) and 3 (blue). We compare for the baseline scale cuts the power spectrum only fits (dashed lines, empty contours) to the one including the tree-level bispectrum (solid lines, filled contours). We find very good agreement with the input cosmologies and confirm the gains for the inclusion of the bispectrum. We tabulate the cosmological and EFT parameters in Tab.~\ref{tab:tab_meandv_freecosmo} for the mean data vector fits for both models.
    }
    \label{fig:freecosmo_compare_M1_M3}
\end{figure}

After validating the unbiased recovery of cosmological parameters from each phase of the \abacus mocks, we fit the mean data vector over all available 12 realizations (six phases and two lines of sight) and divide the covariance matrix by a factor of 6 for models 1 (red) and 3 (blue). The 1D and 2D marginalized posteriors are shown in  Fig.~\ref{fig:freecosmo_compare_M1_M3}. 
We use the baseline scale cuts for each fit and compare the power spectrum only fits (dashed lines, empty contours) to the one including the tree-level bispectrum (solid lines, filled contours). This is a stringent test of our fitting methodology to an (almost) noiseless data vector. We find agreement at the $\approx 1\sigma$-level with the input cosmologies and between both models. 
In Tab.~\ref{tab:tab_meandv_freecosmo} in App.~\ref{sec:tab_meandv_freecosmo} we summarize the results for the fits to the mean data vectors for both models to demonstrate the robustness of the present approach. 

\subsection{Bispectrum mean data vector}\label{sec:res_bk_dv}
\begin{figure}
    \centering
    \includegraphics[width=1\linewidth]{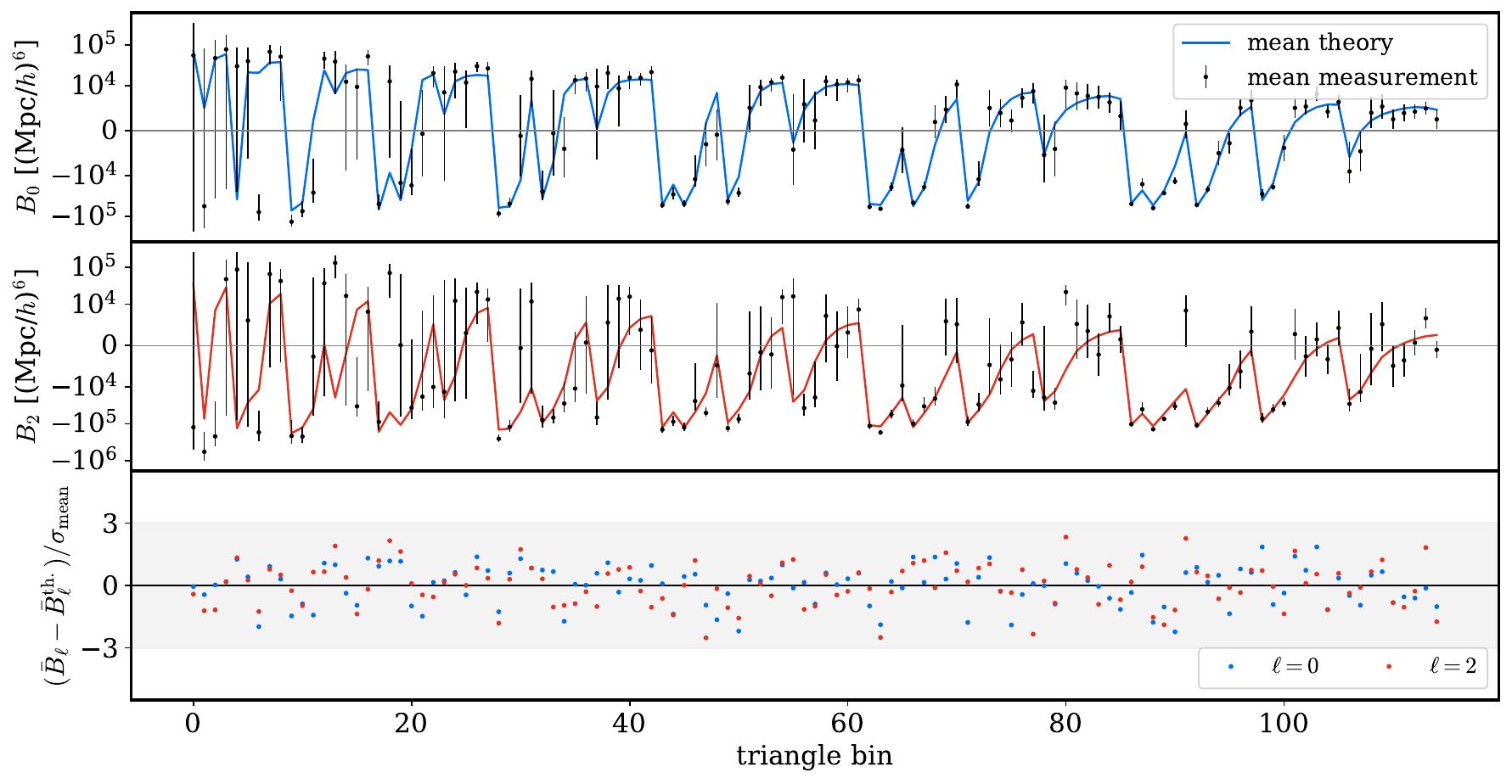}
    \vspace{-0.3in}
    \caption{\textbf{Mean bispectrum measurement.} EFT fit to the mean measured bispectrum  evaluated at the cosmological and bias parameters obtained from the joint fits including the power spectrum (shown in Fig.~\ref{fig:freecosmo_PB_mean}). The monopole ($\ell=0$; top panel) and quadrupole ($\ell=2$; middle panel) measurements and mean best fits are shown with the residual using the error on the mean in the bottom panel. We also show a $3\sigma$ gray error band to guide the eye. 
    The fits use $\kmax=0.1\hMpcinv$ for the bispectrum. 
    We show all 115 triangle bins with sides between $k_{\min}=0.01\hMpcinv$ and $k_{\max}^B=0.1\hMpcinv$, all of which enter the fit. In Fig.~\ref{fig:freecosmo_B_mean_equilater_vs_squeezed} we investigate the performance of our tree-level bispectrum modeling for two triangle configurations as a function of scale cuts. 
    }
    \label{fig:freecosmo_Btree_mean}
\end{figure}

In Fig.~\ref{fig:freecosmo_Btree_mean} we show the monopole and quadrupole of the \Lya forest bispectrum for all triangle configurations with sides between $k_{\min}=0.01\hMpcinv$ and $k_{\max}^B=0.1\hMpcinv$, i.e.,~115 triangle bins and $N_B=230$ data points. We compare the mean measurement to the mean of the best-fit tree-level models of the joint fits with the power spectrum, and show the residuals in units of the error on the mean in the bottom panel. All triangle shapes, from squeezed to equilateral configurations, enter the fit. For a discussion on how to measure the bispectrum in the presence of realistic survey selection function, we refer the reader to Ref.~\cite{deBelsunceSullivanMcDonald2025} for an introduction to a highly informative compression of the \Lya forest bispectra to skew spectra.

\begin{figure}
    \centering
    \includegraphics[width=0.9\linewidth]{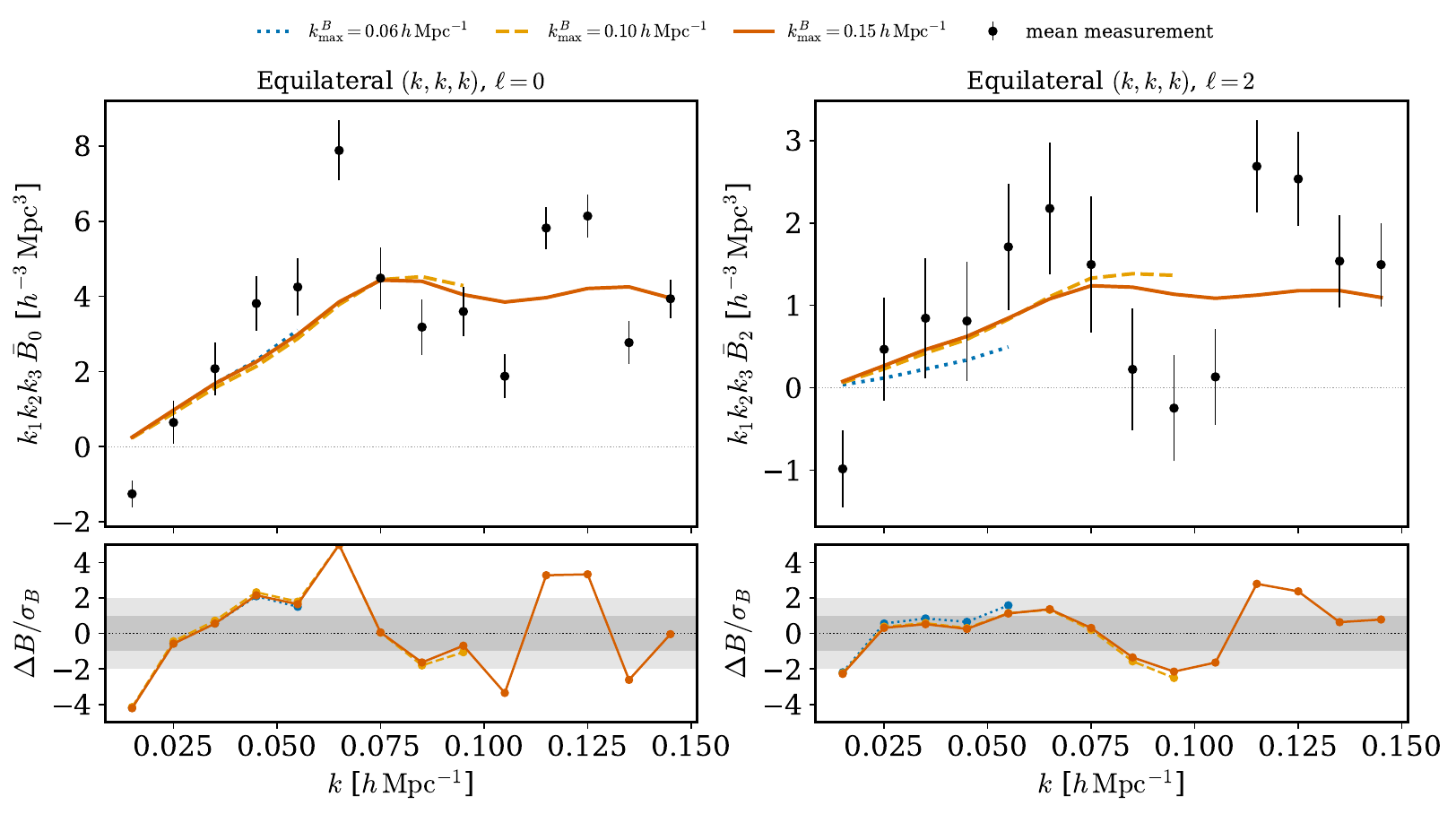}\hfill
    \includegraphics[width=0.9\linewidth]{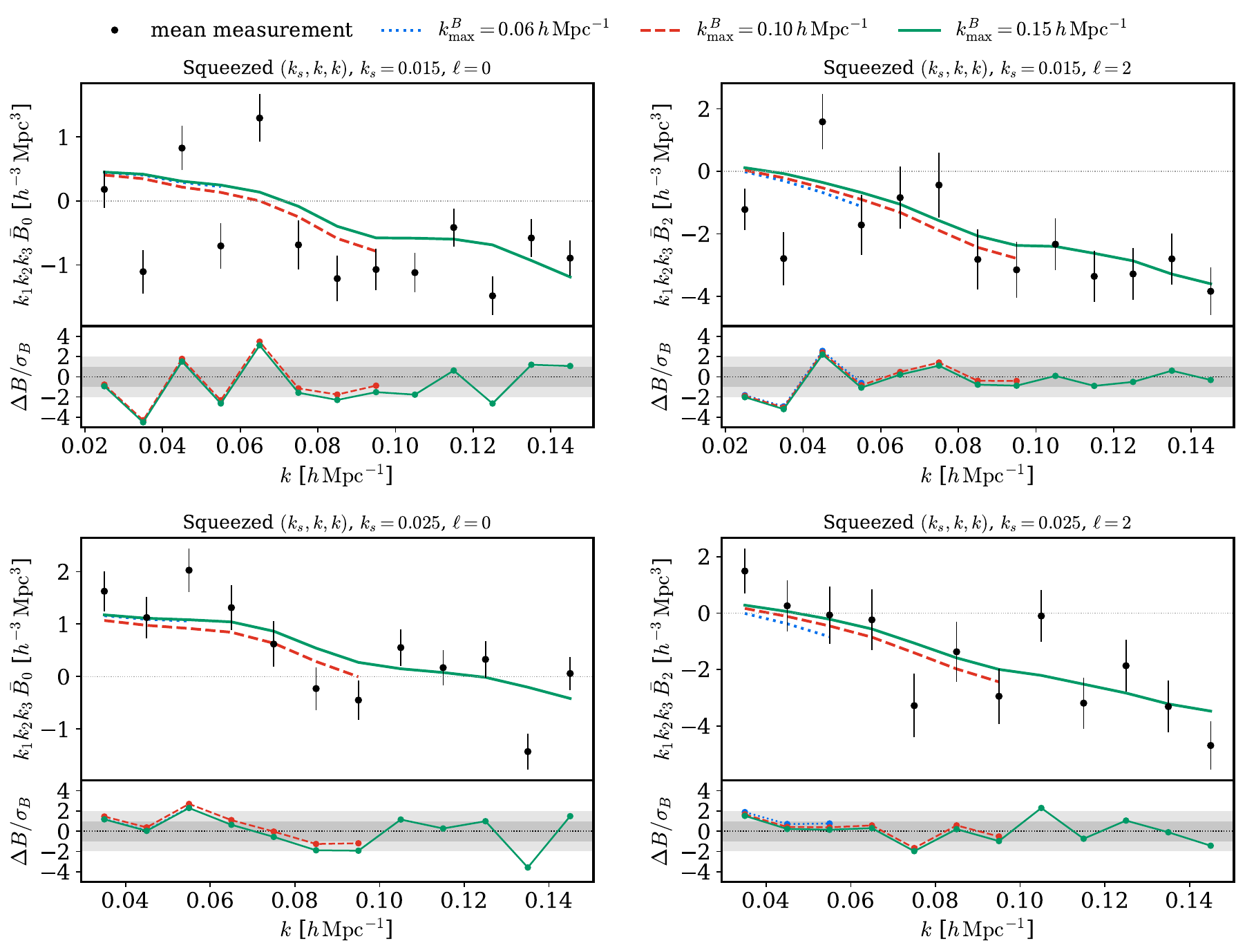}
    \vspace{-0.2in}
    \caption{\textbf{Mean bispectrum measurement for two configurations.} Same as Fig.~\ref{fig:freecosmo_Btree_mean} for the results on the tree-level bispectrum evaluated at the cosmological and bias parameters obtained from the joint fits for the equilateral (top panel) and squeezed configurations (bottom two panels). The monopole ($\ell=0$; left column) and quadrupole ($\ell=2$; right column) measurements and mean best fits for three scale cuts, $k_{\max}^B=0.06,\, 0.10,\, 0.15\,\hMpcinv$, are shown with the residual using the error on the mean in the bottom panel. 
    }
    \label{fig:freecosmo_B_mean_equilater_vs_squeezed}
\end{figure}

Correspondingly we show in Fig.~\ref{fig:freecosmo_B_mean_equilater_vs_squeezed} the mean best-fit tree-level bispectrum for the monopole ($\ell=0$; top panel) and quadrupole ($\ell=2$; center panel) for two triangle configurations:  equilateral (top row) and squeezed (bottom two rows) with the error-normalized residuals for each. We show three  scale cuts with $k_{\max}^B=0.06,\, 0.10,\, 0.15\,\hMpcinv$  to explore the flexibility of the tree-level bispectrum model. For the equilateral monopole model, we find clear deviations at at all scales which is reduced to smaller scales for the quadrupole beyond $\kmax>0.11 \hMpcinv$. Similarly, for very squeezed triangles ($k_s=0.015\hMpcinv$) we find outliers at all scales for the monopole which is somewhat reduced for the quadrupole and less squeezed triangles. We restrict our baseline analyses to $k_{\max}^B=0.10\,\hMpcinv$ and leave a detailed investigation of the covariance matrix and binning potentially causing this discrepancy to future work on simulations.

\subsection{Adding the quasar two and three-point auto-correlation}\label{sec:Pqq_Bqqq}
\begin{figure}
    \centering
    \includegraphics[width=0.49\linewidth]{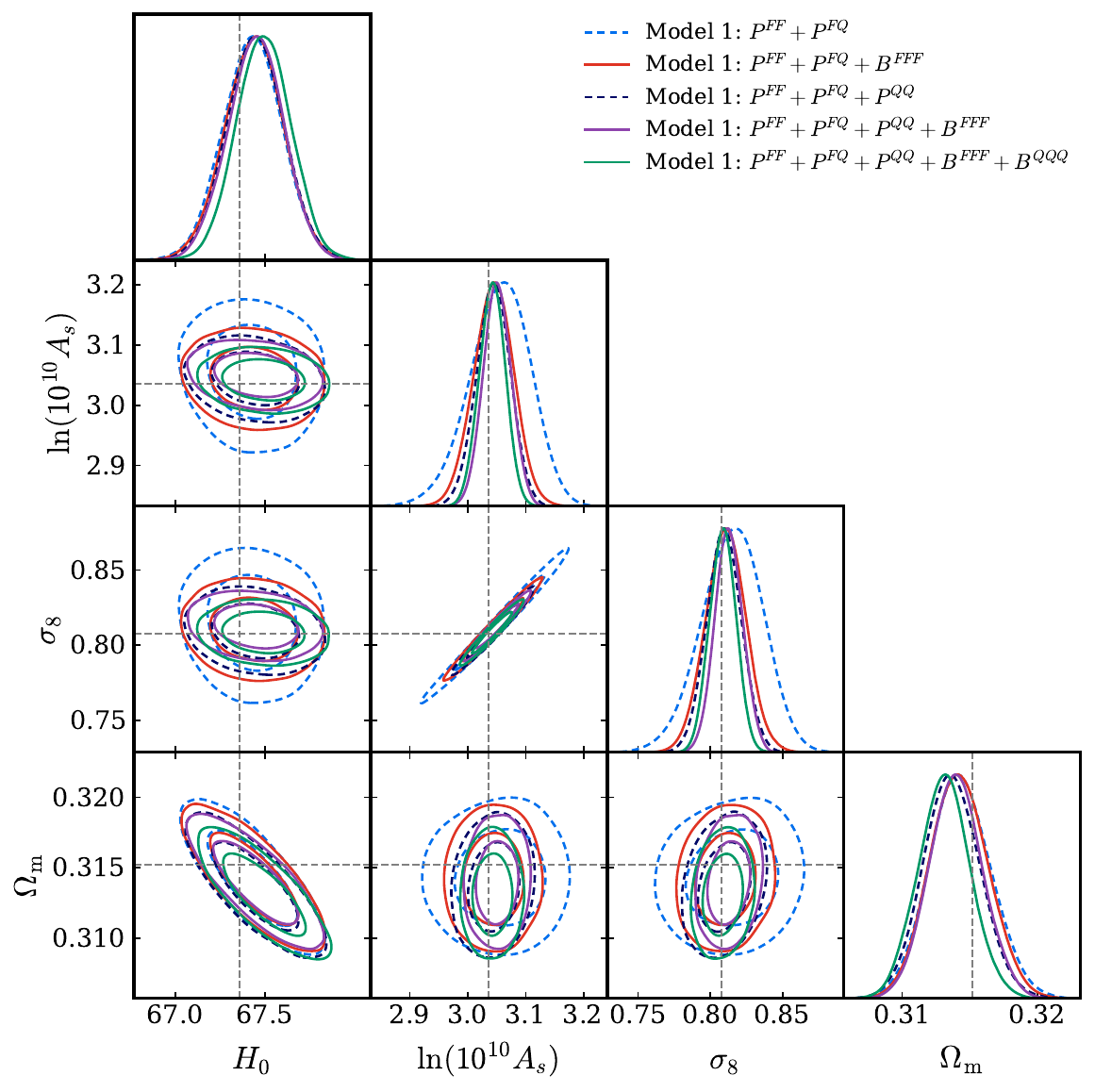}
    \includegraphics[width=0.49\linewidth]{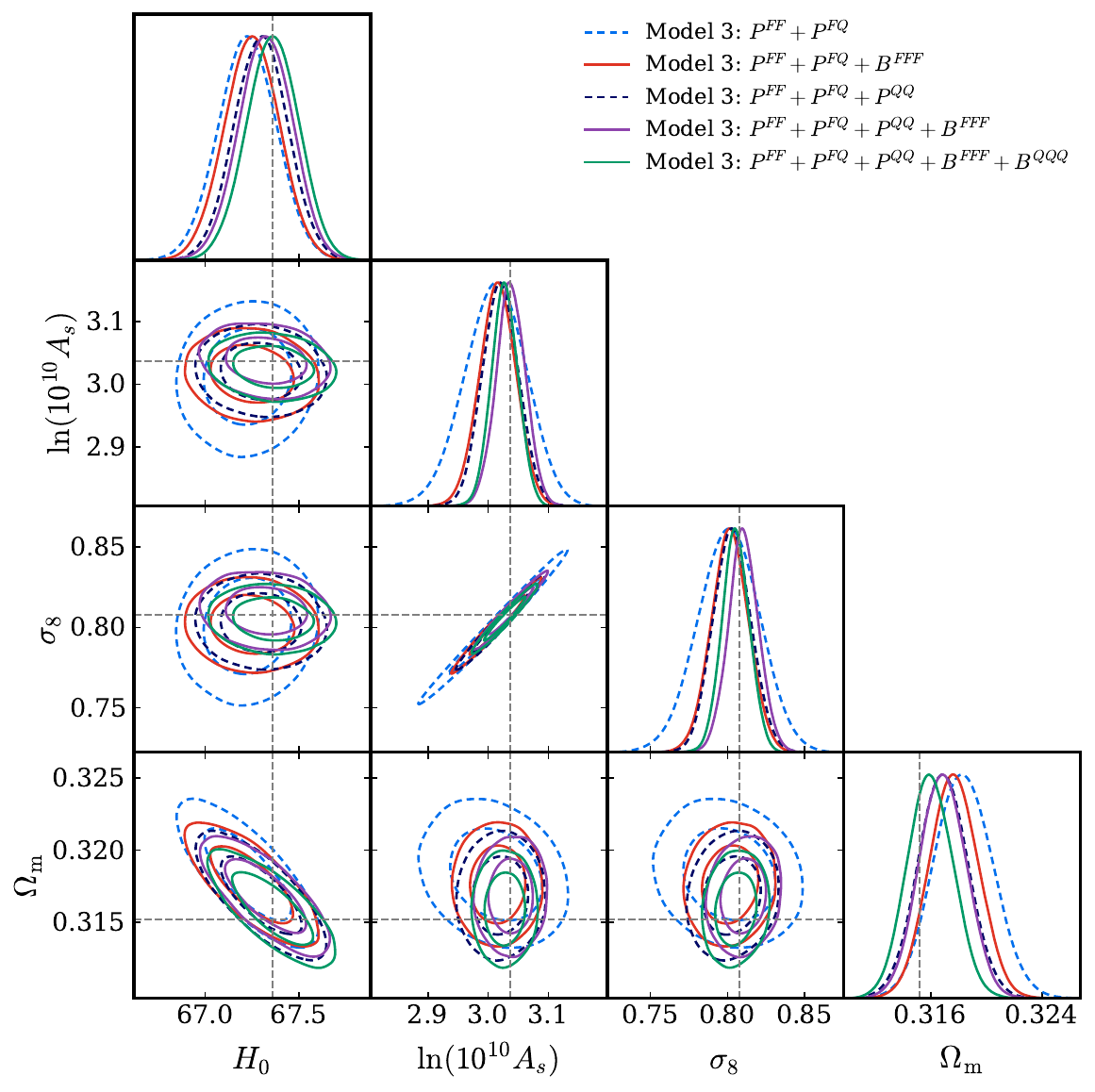}
    \vspace{-0.1in}
    \caption{\textbf{Impact of including the quasar auto-correlation on mean data vector fits.} We investigate the impact of adding different summary statistics to our baseline fits using $P$-only (light blue dashed) and $P+B$ (red solid line) mean data vector analyses for Models 1 (left) and 3 (right panel). First, we add the quasar auto power spectrum (blue dashed) to the $P$ analysis. Second, we add it to the $P+B$ variant ( violet lines). Finally, we add the auto quasar tree-level bispectrum ($QQQ$; green line). For Model 1 for the power spectrum only variant ($FF+FQ+QQ$) the improvements are $43\%$ for $\lnAs$ and $43\%$ for $\sig$. For the ``kitchen sink'' variant including all three power spectra and both bispectrum constraints improve on $\lnAs$, $\sig$, and $\Omega_{\rm m}$ by $\approx 55\%$, $57\%$, and $17\%$, respectively, compared to $P$ alone with constraints on $H_0$ only improving mildly by $\approx 6\%$. We find similar improvements for Model 3 but see that the inclusion of the quasar auto-spectra reduces a discrepancy for $\Omega_{\rm m}$ of $1.5\sigma$ to $0.4\sigma$. This analysis ignores the cross-covariance between both tracers and each of the summary statistics. 
    The results are tabulated in Tab.~\ref{tab:meandv_Pqq_Bqqq}.
    }
    \label{fig:freecosmo_Pqq_Bqqq}
\end{figure}

Current DESI \Lya forest analyses use the combination $FF$ and $FQ$ in the correlation function~\cite{DESI_lya_dr2}. In this section we argue that the inclusion of the quasar auto-power spectrum leads to significant improvements on cosmological parameters and (more optimistically) the quasar auto-bispectrum ($QQQ$) further improves this by including the monopole and quadrupole. We use as scale cuts $k_{\rm max}^{QQ}=0.2 \hMpcinv$ for the power spectrum and for the bispectrum we use $k_{\rm max}^{QQQ}=0.08 \hMpcinv$, following Ref.~\cite{Chudaykin:2025aux}. 
We illustrate the results for both models in Figs.~\ref{fig:freecosmo_compare_M1_M3} and~\ref{fig:freecosmo_Pqq_Bqqq}, and summarize the obtained marginalized posteriors in Tab.~\ref{tab:meandv_Pqq_Bqqq}. We find a progressive improvement when adding additional summary statistics for both models. 

For Model 1 we find the following: First, adding the \Lya forest bispectrum to the standard $FF+FQ$ power-spectrum analysis ($P$) reduces the uncertainty on $\lnAs$ by $\approx 32\%$ and that on $\sig$ by $\approx 33\%$, while only mildly improving the constraint on $\Omega_{\rm m}$ by $\approx 9\%$. The constraint on $H_0$ is virtually unchanged.
Relative to $P$, adding $P_{QQ}$ (third column in Tab.~\ref{tab:meandv_Pqq_Bqqq}) reduces the uncertainties on $\lnAs$ and $\sig$ by $\approx 44\%$ and $\approx 43\%$, respectively, and improves the $\Omega_{\rm m}$ constraint by $\approx 9\%$. Thus, $P+P_{QQ}$ already yields tighter amplitude constraints than $P+B$, despite not requiring the \Lya forest bispectrum.
The addition of $B_{FFF}$ to $P+P_{QQ}$ provides a further $\approx 17\%$ improvement on $\lnAs$ and $\approx 20\%$ on $\sig$, with a more modest $\approx 5\%$ gain on $\Omega_{\rm m}$. Finally, adding $B_{QQQ}$ reduces the remaining uncertainties by a further $\approx 4\%$, $5\%$, and $5\%$ for $\lnAs$, $\sig$, and $\Omega_{\rm m}$, respectively. 

The same qualitative pattern is obtained for Model 3. Adding $B_{FFF}$ to $P$ improves the constraints on $\lnAs$, $\sig$, and $\Omega_{\rm m}$ by $\approx 39\%$, $40\%$, and $14\%$, respectively. Adding $P_{QQ}$ to $P$ yields corresponding improvements of $\approx 41\%$, $40\%$, and $14\%$. The corresponding gains from adding $B_{FFF}$ to $P+P_{QQ}$ are $\approx 17\%$, $18\%$, and $6\%$, while the subsequent addition of $B_{QQQ}$ improves $\lnAs$ and $\sig$ by a further $\approx 12\%$ and $10\%$, respectively. The $\Omega_{\rm m}$ width is unchanged by the addition of $B_{QQQ}$ for Model 3 whilst it pulls the mean value closer to the true one. We illustrate the results for both models in Figs.~\ref{fig:freecosmo_compare_M1_M3} and~\ref{fig:freecosmo_Pqq_Bqqq}, and summarize the obtained marginalized posteriors in Tab.~\ref{tab:meandv_Pqq_Bqqq}. We find a progressive improvement when adding additional summary statistics for both models.

The full ``kitchen sink analysis'' (last column in Tab.~\ref{tab:meandv_Pqq_Bqqq})  improves the constraints on $\lnAs$, $\sig$, and $\Omega_{\rm m}$ by $\approx 55\%$, $57\%$, and $17\%$, respectively, compared to $P$ alone with constraints on $H_0$ only improving mildly by $\approx 6\%$. For Model 3, the corresponding improvements are $\approx 57\%$, $56\%$, and $19\%$. In particular, for $\Omega_{\rm M}$ we find that including the quasar auto information reduces a $ 1.5 \sigma$ bias to $0.4\sigma$. 
Overall, these results motivate including $P_{QQ}$ as a priority extension of the standard $FF+FQ$ analysis, and eventually $B_{QQQ}$, following Ref.~\cite{Chudaykin:2025aux}.

\begin{table*}
    \centering
    \setlength{\tabcolsep}{12pt}
    \renewcommand{\arraystretch}{1.11}
    \begin{tabular}{@{}lccccc@{}}
    \hline
    Parameter & $P$ & $P+B$ & $P+P_{QQ}$ & $P+P_{QQ}+B$ & $P+P_{QQ}+B+B_{QQQ}$ \\
    \hline
    \textbf{Model 1} &&&&& \\
    $H_0$ & $67.43\pm0.16$ & $67.44\pm0.16$ & $67.45\pm0.16$ & $67.45\pm0.15$ & $67.49\pm0.15$ \\
    $\ln(10^{10}A_s)$ & $3.055^{+0.056}_{-0.047}$ & $3.044\pm0.035$ & $3.044\pm0.029$ & $3.051\pm0.024$ & $3.042\pm0.023$ \\
    $\sigma_8$ & $0.815^{+0.022}_{-0.020}$ & $0.811\pm0.014$ & $0.809\pm0.012$ & $0.8127\pm0.0096$ & $0.8085\pm0.0091$ \\
    $\Omega_{\rm m}$ & $0.3143\pm0.0023$ & $0.3142\pm0.0021$ & $0.3137\pm0.0021$ & $0.3140\pm0.0020$ & $0.3131\pm0.0019$ \\
    \hline
    \textbf{Model 3} &&&&& \\
    $H_0$ & $67.22\pm0.15$ & $67.25\pm0.15$ & $67.30\pm0.14$ & $67.32\pm0.14$ & $67.36\pm0.14$ \\
    $\ln(10^{10}A_s)$ & $3.012\pm0.051$ & $3.016\pm0.031$ & $3.021\pm0.030$ & $3.038\pm0.025$ & $3.027\pm0.022$ \\
    $\sigma_8$ & $0.801\pm0.020$ & $0.802\pm0.012$ & $0.803\pm0.012$ & $0.8103\pm0.0099$ & $0.8053\pm0.0089$ \\
    $\Omega_{\rm m}$ & $0.3183\pm0.0021$ & $0.3176\pm0.0018$ & $0.3169\pm0.0018$ & $0.3168\pm0.0017$ & $0.3159\pm0.0017$ \\
    \hline\hline
    \end{tabular}
    \caption{\textbf{Cosmological constraints from mean-data-vector fits.}
    One-dimensional marginalized $68\%$ C.L. constraints from fits to the mean data vector of the 12 Model 1 (top) and Model 3 (bottom block) mocks, shown in Fig.~\ref{fig:freecosmo_compare_M1_M3} for the first two columns and for the remaining columns in Fig.~\ref{fig:freecosmo_Pqq_Bqqq}. The first two rows list sampled parameters, while the two bottom rows ($\sigma_8$ and $\Omega_{\rm m}$) quote derived parameters. All fits use the baseline scale cuts and progressively add summary statistics. 
    $P$ denotes the \Lya forest auto- and cross-power spectra, 
    $FF$ and $FQ$, with $k_{\rm max}^{FF\,(FQ)}=0.6\,(0.3)\,h\,{\rm Mpc}^{-1}$. 
    $B$ denotes the \Lya forest auto-bispectrum, $FFF$, with $k_{\rm max}^{FFF}=0.1\,h\,{\rm Mpc}^{-1}$. 
    $P+P_{QQ}$ adds the quasar auto-power spectrum with $k_{\rm max}^{QQ}=0.2\,h\,{\rm Mpc}^{-1}$; 
    $P+P_{QQ}+B$ additionally includes $FFF$; and 
    $P+P_{QQ}+B+B_{QQQ}$ further adds the quasar auto-bispectrum with $k_{\rm max}^{QQQ}=0.08\,h\,{\rm Mpc}^{-1}$. 
    We fit all power spectra as $P(k,\mu)$ and include the monopole and quadrupole of both bispectra.
    }  
    \label{tab:meandv_Pqq_Bqqq}
\end{table*}

\section{Summary \& Conclusion}\label{sec:conclusion}

The \Lya forest is a pristine tracer of the large-scale structure of the Universe at high redshift ($2 \leq z \leq 4$) down to Mpc-scales and below. It currently provides constraints at the one percent-level on the expansion history during matter domination ($z\approx 2.33$) which is highly complementary to low-redshift galaxy clustering measurements from DESI. The \Lya forest probes a regime that is inaccessible to current and near-future galaxy clustering surveys, making it crucial to maximize the amount of information that we extract from the data to constrain cosmological physics. 

In this work, we have developed a joint theoretical model within the effective field theory (EFT) framework for the one-loop power spectrum and tree-level bispectrum. Given a smaller set of symmetries compared to galaxies and quasars, the \Lya forest requires additional parameters along the line-of-sight.
We demonstrate our methodology on large-volume \Lya forest mocks painted on top of the \textsc{AbacusSummit} $N$-body simulation suite. We first validate the model for the key Fourier-space data combination used by the DESI collaboration: the \Lya forest auto-spectrum, $FF$, and the forest--quasar cross-spectrum, $FQ$.

For our baseline analysis, we use $k_\mathrm{max}^{FF\,(FQ)}=0.6\,(0.3)\,h\,\mathrm{Mpc}^{-1}$ for the power spectra and $k_\mathrm{max}^{FFF}=0.1\,h\,\mathrm{Mpc}^{-1}$ for the \Lya forest bispectrum. For fits to individual mock realizations, we recover all cosmological parameters within $\approx1\sigma$ already with the power spectrum alone. In particular, the single-realization constraints on the cosmological parameters are for the power spectrum alone ($P$) and when including the bispectrum ($P+B$) given by
\begin{subequations}
\begin{align}
H_0 &= 67.48 \pm 0.38\,\mathrm{km}\,\mathrm{s}^{-1}\,\mathrm{Mpc}^{-1},
&
\ln(10^{10}A_s) &= 3.01^{+0.11}_{-0.085},
&
\sigma_8 &= 0.806^{+0.042}_{-0.036},
&& (P)\,, \\
H_0 &= 67.61 \pm 0.38\,\mathrm{km}\,\mathrm{s}^{-1}\,\mathrm{Mpc}^{-1},
&
\ln(10^{10}A_s) &= 3.024 \pm 0.071,
&
\sigma_8 &= 0.809 \pm 0.028,
&& (P+B)\,.
\end{align}
\end{subequations}
This corresponds to a precision on the Hubble constant of $\approx0.6\%$ for the power spectrum alone with an identical result when adding the \Lya forest bispectrum. The baseline bispectrum fit improves the constraints on $\lnAs$ and $\sig$ by $27\%$ and $28\%$, respectively, relative to the power-spectrum-only fit. The agreement of the recovered parameters between the power-spectrum-only and both joint fits provides an important validation of the joint model.
For the same fits we additionally improve on the bias parameters by adding the bispectrum. This results in improvements on $b_2$ by 72\%, on $b_{\mathcal{G}_2}$ by 67\%, on $b_{(KK)_\parallel}$ by 52\% and on the linear bias parameter, $b_1$, by 32\%. Except for $b_\eta$ we find strong improvements in all quoted constraints. 
We find similar yet slightly weaker improvements for the remaining quadratic bias parameters and virtually none for the cubic bias parameters and counterterms matching baseline expectations. 

Further we investigate the impact of including  the quasar power spectrum ($QQ$) and bispectrum ($QQQ$). First, we add the \Lya forest bispectrum to the baseline $FF+FQ$ power-spectrum analysis which improves the constraints on $\lnAs$ and $\sig$ by $\approx32\%$ and $\approx33\%$, respectively, while the $\Omega_{\rm m}$ constraint improves by $\approx9\%$ and the $H_0$ constraint is virtually unchanged. However, adding $P_{QQ}$ to the baseline power-spectrum analysis adds additional information and reduces the uncertainties on $\lnAs$ and $\sig$ by $\approx44\%$ and $\approx43\%$, respectively. The constraints on $\Omega_{\rm m}$ improve by a more modest $\approx9\%$. The resulting power spectrum only data vector, $P+P_{QQ}$ already provides tighter amplitude constraints than $P+B$, without requiring the complexities coming from the \Lya forest bispectrum.

Now adding to $P+P_{QQ}$ the \Lya forest bispectrum yields a further $\approx17\%$ improvement on $\lnAs$ and $\approx20\%$ on $\sig$, together with a $\approx5\%$ gain on $\Omega_{\rm m}$. Adding $B_{QQQ}$ on top of this combination provides a small improvement of $\approx4\%$, $5\%$, and $5\%$ for $\lnAs$, $\sig$, and $\Omega_{\rm m}$, respectively. The full combination of $FF$, $FQ$, $P_{QQ}$, $FFF$, and $B_{QQQ}$ improves the constraints on $\lnAs$, $\sig$, and $\Omega_{\rm m}$ by $\approx55\%$, $57\%$, and $17\%$, respectively, compared to the baseline power-spectrum-only analysis. The $H_0$ constraint improves only mildly, by $\approx6\%$.

\textit{In summary, these results highlight the promise of the quasar auto-power spectrum as a particularly important extension of the standard $FF+FQ$ analysis.} It provides more information on the amplitude parameters than the \Lya forest bispectrum alone from the inclusion of all multipoles by measuring $f\sig$ directly in contrast to the \Lya forest auto spectrum where $b_\eta$ is an additional free parameter.
The quasar bispectrum provides an additional, smaller gain once the other observables are included and is a natural subsequent extension~\cite{Chudaykin:2025aux}. Together, these results are very encouraging and highlight the promise for discriminating between dynamical dark energy models at high redshift using the \Lya forest~\cite{Weiner:2026sfm}.

These tests should be treated as idealized, high-precision tests for the EFT models as we are using fully sampled boxes resulting in unrealistically small error bars and also fitting mean data vectors. Additionally, the FGPA procedure should be treated as an approximation to the \Lya forest flux that introduces a higher degree of non-linearity as the effects of pressure smoothing (removing small-scale structure) are ignored. We emphasize that we demonstrate the unbiased recovery of cosmological parameters for the one-loop power spectrum and tree-level bispectrum in a highly idealized scenario. So far, we have ignored  a plethora of data-specific effects: the instrument response and pixel noise, continuum fitting, finite forest lengths and gaps in the spectra, sky residuals, an anisotropic mask and selection function, the redshift evolution of the spectra, redshift errors for quasars, and high column density systems and metals. These models are now readily applicable to mocks including DESI-specific observational effects as implemented in the \texttt{CoLoRe-2LPT} mocks~\cite{Bernal:2026klr}. In particular, to deal with the pathological window function of the \Lya forest, shifted skew spectra -- a compression of the bispectrum by correlating the product of two copies of the field of which one is displaced in the radial direction, with the original one -- can be evaluated with the same theory Model \cite{deBelsunceSullivanMcDonald2025}.

To improve upon this work, we suggest developing the one-loop bispectrum for the \Lya forest, implementing an angular-dependent scale cut due to the velocity field driving the breakdown of EFT at small scales, including the cross-covariance between correlated fields as well as cross-bispectra between flux and point tracers ($FFQ$ and $FQQ$) which could be used to test the equivalence principle as the \Lya forest traces baryons and quasars (mostly) trace dark matter (see, e.g.,~\cite{Peloso:2013zw,Kehagias:2013yd}). In the present analysis we do not remove self-skewer contributions~\cite[see, e.g.][]{deBelsunce:2024knf} but the present EFT framework allows us to model the one-dimensional power spectrum (P1D;~\cite{Ivanov:2024jtl}) and bispectrum (B1D;~\cite{delaCruz:2024cai}). The same model can readily be extended to include cross-correlations with lensing of the cosmic microwave background to further break the degeneracy of the growth rate with the (unknown) velocity
gradient bias through a remapping of the bias parameters~\cite{Ivanov:2026dvl, deBelsunce:2025qku} and can be compared to hybrid EFT models that compute the displacements using $N$-body simulations~\cite{deBelsunce:2025gci}. 

It will also be interesting to extend our analysis to primordial non-Gaussianity
in the bispectrum, which is especially relevant now as the galaxy clustering
data produces constraints 
competitive with those obtained from the CMB~\cite{Cabass:2022epm,Cabass:2022wjy,Chaussidon:2024qni,Chudaykin:2025vdh}. This extension, however, will require an accurate treatment of the bispectrum window function. 

We leave the detailed exploration of these research avenues to future work.

\acknowledgments
We thank Pat McDonald
and the DESI \Lya forest working group
for fruitful discussions. 
RdB acknowledges productive conversations with Arnaud de Mattia and Lehman Garrison on CUDA kernel optimizations on the GPU. 

JMS acknowledges that support for this work was provided by The Brinson Foundation through a Brinson Prize.
MMI is supported by the Office of High Energy Physics of U.S. Department of Energy under grant Contract Number DE-SC0012567.
This work is supported by the National Science Foundation under Cooperative Agreement PHY-2019786 (The NSF AI Institute for Artificial Intelligence and Fundamental Interactions, \url{http://iaifi.org/}). 
This research used resources of the National Energy Research Scientific Computing Center (NERSC), a U.S. Department of Energy Office of Science User Facility operated under Contract No. DE-AC02-05CH11231. Parts of the computations in this paper were run on the FASRC Cannon cluster supported by the FAS Division of Science Research Computing Group at Harvard University.

\textit{AI usage statement:} This work was entirely conceptualized and written by the authors. Claude (Opus 5.5 and Fable 5.1 through Anthropic's AI for Science program) was used, in parts, to perform numerical computations and for code optimization in \texttt{python} and to write \texttt{CUDA} kernels as well as to edit the manuscript for clarity. The authors have checked and validated all the results, taking full responsibility for all results presented herein.

\bibliographystyle{JHEP}
\bibliography{bib}

\appendix

\section{Impact of scale cuts on cosmological analysis}\label{sec:cosmo_scale_cuts}

\begin{figure}
    \centering
    \includegraphics[width=.8\linewidth]{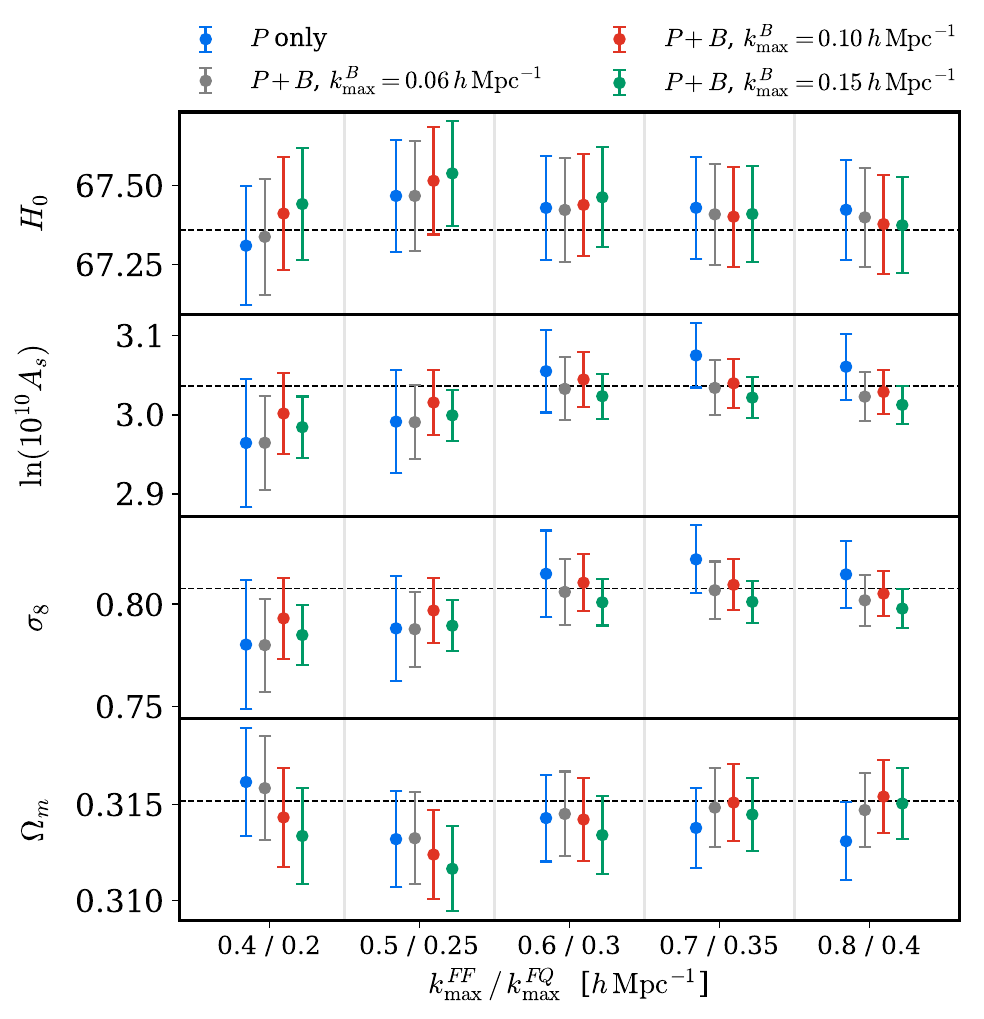}
    \vspace{-0.2in}
    \caption{\textbf{Determination of scale cuts from mean data vector fits:} Fits to the mean data vector over all available 12 realizations (six phases and two lines of sight) and dividing the covariance matrix by a factor of 12 for Model 1 for a range of scale cuts. We investigate pairs of scale cuts; namely for the $FF$ and $FQ$ power spectra of $k_{\rm max} = (0.4,\, 0.2)\, (0.5,\, 0.25)\, (0.6,\, 0.3)\, (0.7,\, 0.35)\, (0.8,\, 0.4) \hMpcinv$ with three bispectrum scale cuts of $k_{\rm max}^B=0.06,\, 0.10,\, 0.15 \hMpcinv$. We find excellent agreement with the input cosmologies and obtained increasingly tight constraints when using smaller scales. 
    }
    \label{fig:freecosmo_kmax}
\end{figure}

In this appendix we show the fits to the mean data vector of the 12 mocks (six phases and two lines of sight) with free cosmology and  use these to determine our scale cuts 
in Fig.~\ref{fig:freecosmo_kmax} using a drift plot. We investigate the impact of varying the maximum wavenumber used in the power spectrum and bispectrum fits and its impact on the recovery of the cosmological parameters. For each set of scale cuts we show the 1D marginalized posterior of four combinations: power spectrum only (blue), three bispectrum scale cuts with increasing range from gray to red to green. Our baseline results are given in red. We observe clear improvement in $\sig$ and $\lnAs$ when including smaller scales of the bispectrum. Whilst the results are remarkably consistent across the different $\kmax$ values, we choose a conservative scale cut of $k_\mathrm{max}^{FF\,(FQ)}=0.6\,(0.3)\,h\,\mathrm{Mpc}^{-1}$ for the power spectrum and $k_\mathrm{max}^{FFF}=0.1\,h\,\mathrm{Mpc}^{-1}$ for the bispectrum. For the bispectrum we perform additional sanity checks, shown in Fig.~\ref{fig:freecosmo_B_mean_equilater_vs_squeezed}, to investigate the agreement between the tree-level bispectrum theory modeling for the equilateral and squeezed configurations.

\section{\textsc{AbacusSummit} realizations}\label{sec:app_phases}

\begin{figure}
    \centering
    \includegraphics[width=\linewidth]{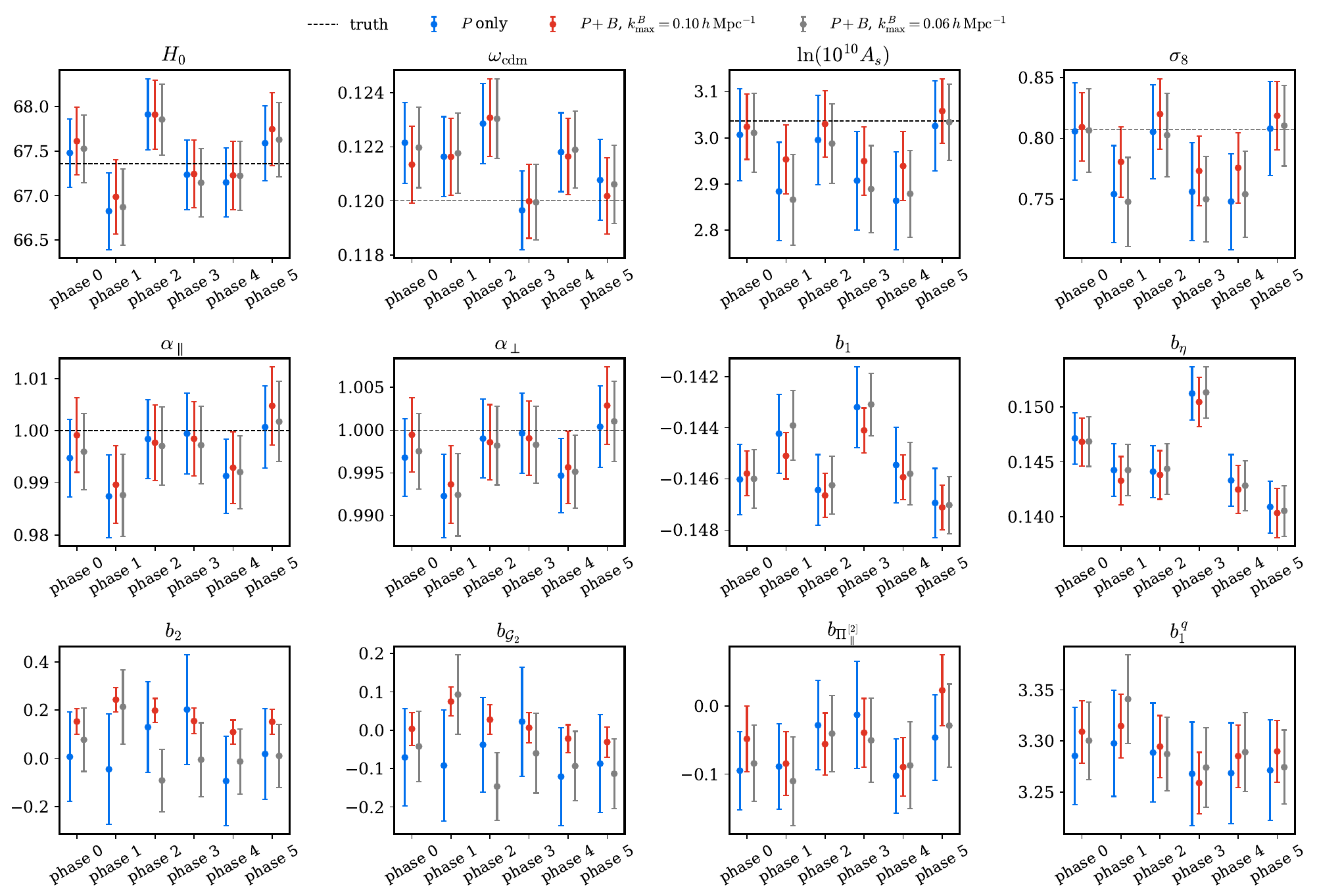}
    \vspace{-0.2in}
    \caption{\textbf{Overview of free-cosmology constraints.} Same as Fig.~\ref{fig:freecosmo_PB} for each phase for model 1 showing the 1D marginalized posteriors. We recover the sampled cosmological parameters $H_0,\, \lnAs,\, \omega_{\rm cdm}$ at the $\approx1-1.5\sigma$ level. The fits are obtained by fitting summary statistics from \abacus simulation for model 1 for each of the ``phases'' which corresponds to different noise realizations along the $\hat z$-axis. We compare three sets of fits using only the power spectrum ($FF$ and $FQ$) shown in blue and two variants when adding the tree-level bispectrum with two scale cuts $k_{\rm max}^B=0.06,\, 0.10\,\hMpcinv$ shown in gray and red, respectively. The input values are shown as gray dashed lines. The BAO scaling parameters (Sec.~\ref{sec:EFT_AP}) are derived parameters. 
    }
    \label{fig:freecosmo_PB_1D}
\end{figure}

\begin{figure}
    \centering
    \includegraphics[width=\linewidth]{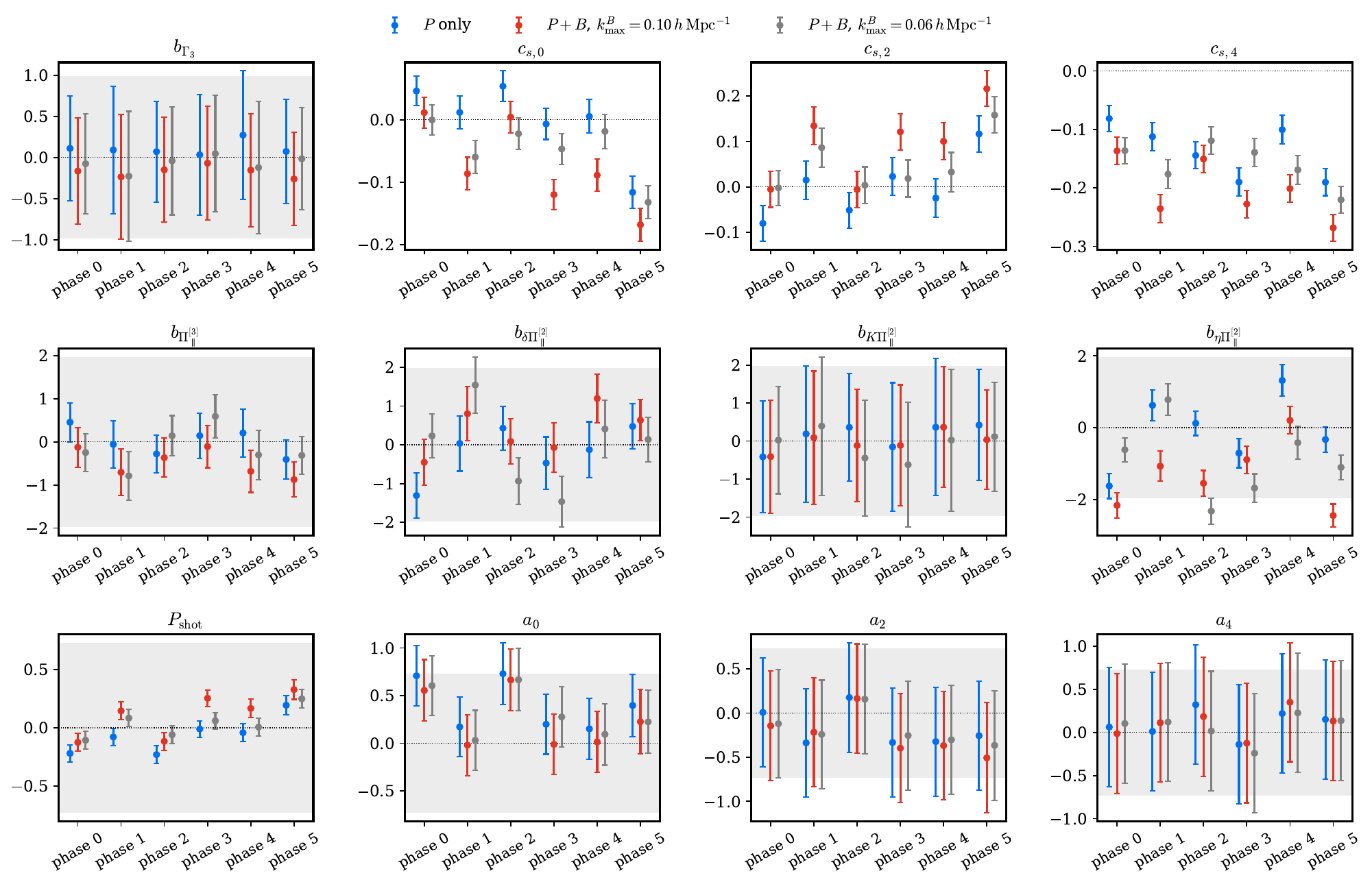}
    \vspace{-0.2in}
    \caption{\textbf{Overview of free-cosmology constraints: \Lya forest auto.} Same as Fig.~\ref{fig:freecosmo_PB_1D} for the higher-order bias parameters of the power spectrum.
    }
    \label{fig:freecosmo_PB_1D_partII}
\end{figure}

\begin{figure}
    \centering
    \includegraphics[width=\linewidth]{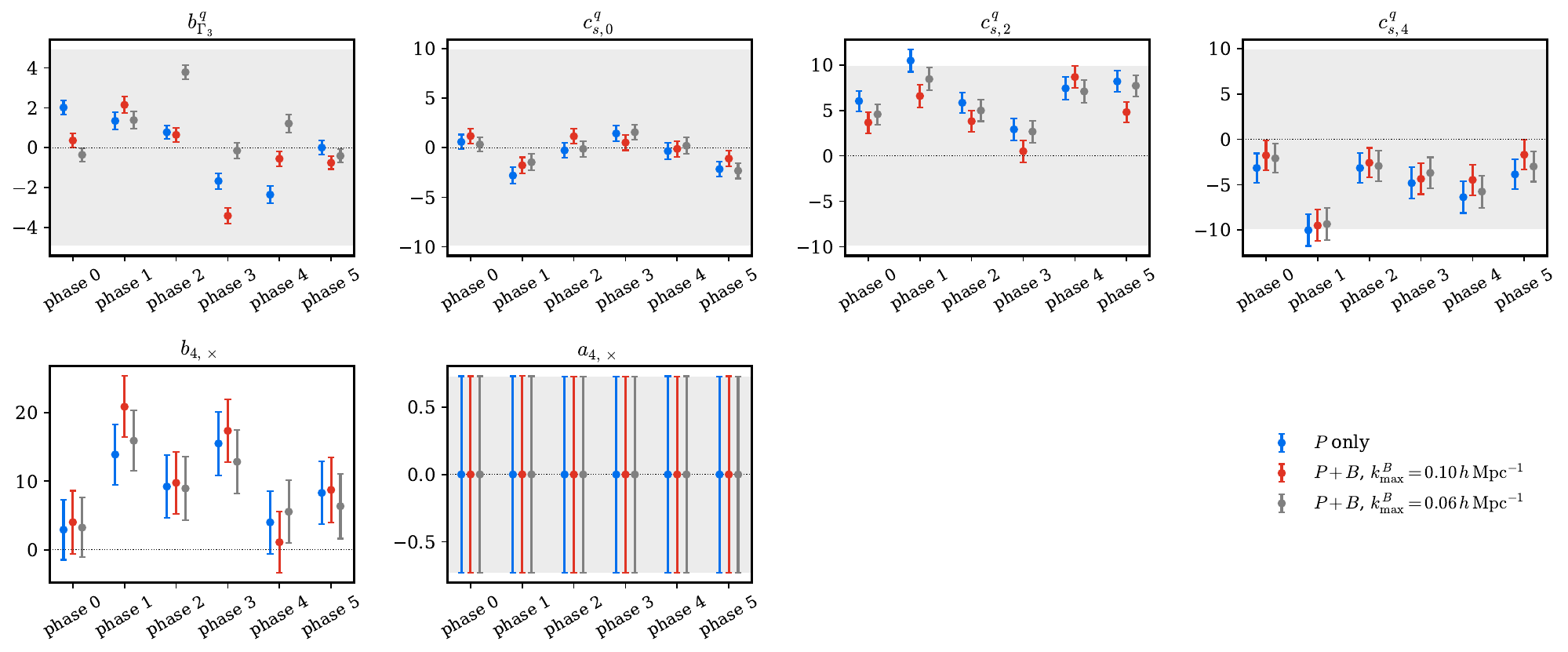}
    \vspace{-0.2in}
    \caption{\textbf{Overview of free-cosmology constraints: \Lya forest -- quasar cross-correlation} Same as Fig.~\ref{fig:freecosmo_PB_1D} for the higher-order bias parameters relevant to the cross-spectrum
    }
    \label{fig:freecosmo_PB_1D_partIII}
\end{figure}

\begin{figure}
    \centering
    \includegraphics[width=\linewidth]{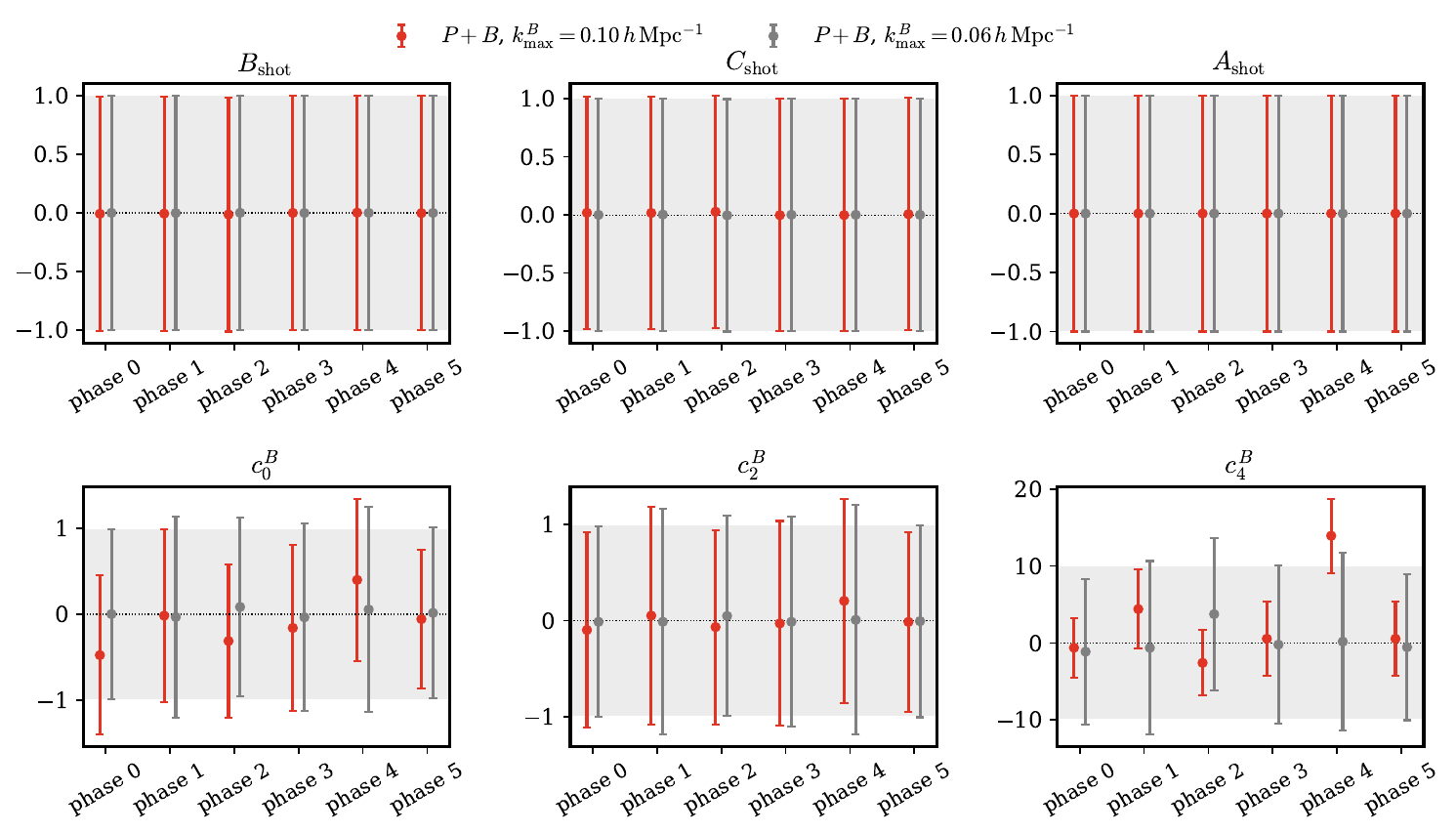}
    \vspace{-0.2in}
    \caption{\textbf{Overview of free-cosmology constraints: \Lya forest bispectrum} Same as Fig.~\ref{fig:freecosmo_PB_1D} for the nuisance parameters of the tree-level \Lya forest bispectrum. 
    }
    \label{fig:freecosmo_PB_1D_partIV}
\end{figure}

We investigate the evolution of the cosmological and bias parameters as a function of the six realizations (phases) for Model 1 on the \textsc{AbacusSummit} simulation. In Figs.~\ref{fig:freecosmo_PB_1D}--~\ref{fig:freecosmo_PB_1D_partIV} we show the recovered cosmological and bias parameters for each realization of the \Lya forest mocks along the $\hat z$-axis. The first set of panels shows the cosmological and leading bias parameters for the \Lya forest and quasars from the power spectrum. The second set of panels shows the higher order bias parameters as well as stochastic and counterterms. The third set of panels is for the quasar and cross-correlation parameters. The fourth panel shows the 1D marginalized constraints for the bispectrum parameters.  We find consistent results among the three different configurations with scatter at the 1$\sigma$-level for the recovery of the cosmological parameters. Including the bispectrum yields improvements of $\approx 28\%$ on the amplitude of matter clustering ($\sig$) and of $\approx  26\%$ on the amplitude of the primordial scalar power spectrum ($\lnAs$). We find prior dominated results for most of the higher-order bias and the bispectrum parameters for which we advocate to use simulation-based priors~\cite[see, e.g.,][]{Ivanov:2024xgb}. 

\section{Constraints from mean data vector with free cosmology}\label{sec:tab_meandv_freecosmo}
In Tab.~\ref{tab:tab_meandv_freecosmo} we summarize the results for the cosmological and EFT nuisance parameters associated with the 1D and 2D marginalized posteriors shown in Fig.~\ref{fig:freecosmo_compare_M1_M3}. For each model (1: first two columns; 3: last two columns) we compare the $P$ and $P+B$ fits. The mean data vector for the power spectrum are shown in Fig.~\ref{fig:freecosmo_PB_mean} and for the bispectrum in Fig.~\ref{fig:freecosmo_Btree_mean} for Model 1. It is interesting to note that the higher-order bias parameters as well as the ones for the bispectrum are poorly constrained and, for some, even prior dominated. The resulting $\chi^2$ values and degrees of freedom, $N$, are quoted for the power spectrum and bispectrum separately as we co-add the likelihoods without any cross-covariance. The power spectrum (bispectrum) has 1772 (230) degrees of freedom and we find that the reduced $\chi^2$ is much smaller than one given that we use a rescaled covariance matrix. 

\begin{table}
    \centering
{
\setlength{\tabcolsep}{15pt}
\renewcommand{\arraystretch}{1.15}
\begin{tabular}{@{}lcccc@{}}
\hline
Parameter & \multicolumn{2}{c}{Model 1} & \multicolumn{2}{c}{Model 3} \\
\cline{2-3}\cline{4-5}
          & $P$ only & $P+B$ & $P$ only & $P+B$ \\
\hline
\multicolumn{5}{@{}l}{\textbf{Sampled and derived parameters}} \\
$H_0$ & $+67.430 \pm 0.164$ & $+67.439 \pm 0.162$ & $+67.225 \pm 0.154$ & $+67.250 \pm 0.146$ \\
$\omega_{\rm cdm}$ & $+0.1199 \pm 0.0007$ & $+0.1199 \pm 0.0006$ & $+0.1208 \pm 0.0006$ & $+0.1206 \pm 0.0005$ \\
$\ln(10^{10}A_s)$ & $+3.0549 \pm 0.0522$ & $+3.0445 \pm 0.0348$ & $+3.0116 \pm 0.0507$ & $+3.0157 \pm 0.0305$ \\
$\sigma_8$ & $+0.8149 \pm 0.0212$ & $+0.8105 \pm 0.0140$ & $+0.8008 \pm 0.0198$ & $+0.8016 \pm 0.0121$ \\
$\Omega_m$ & $+0.3143 \pm 0.0023$ & $+0.3142 \pm 0.0021$ & $+0.3183 \pm 0.0021$ & $+0.3176 \pm 0.0018$ \\
$\qpar$ & $+1.0013 \pm 0.0033$ & $+1.0015 \pm 0.0032$ & $+0.9954 \pm 0.0031$ & $+0.9964 \pm 0.0026$ \\
$\qperp$ & $+1.0008 \pm 0.0020$ & $+1.0009 \pm 0.0019$ & $+0.9972 \pm 0.0019$ & $+0.9978 \pm 0.0016$ \\
$b_1$ & $-0.1459 \pm 0.0009$ & $-0.1460 \pm 0.0004$ & $-0.1305 \pm 0.0006$ & $-0.1306 \pm 0.0004$ \\
$b_\eta$ & $+0.1461 \pm 0.0013$ & $+0.1462 \pm 0.0012$ & $+0.2816 \pm 0.0014$ & $+0.2816 \pm 0.0013$ \\
$b_2$ & $+0.298 \pm 0.141$ & $+0.1751 \pm 0.0239$ & $+0.2450 \pm 0.0988$ & $+0.2178 \pm 0.0302$ \\
$b_{\mathcal{G}_2}$ & $+0.0423 \pm 0.0979$ & $+0.0020 \pm 0.0180$ & $+0.0529 \pm 0.0704$ & $+0.0392 \pm 0.0206$ \\
$b_{\eta^2}$ & $-0.0869 \pm 0.0298$ & $-0.0608 \pm 0.0225$ & $-0.0445 \pm 0.0342$ & $-0.0291 \pm 0.0280$ \\
$b_{\delta\eta}$ & $+0.1002 \pm 0.0970$ & $+0.1245 \pm 0.0388$ & $+0.2757 \pm 0.0696$ & $+0.3023 \pm 0.0429$ \\
$b_{(KK)_\parallel}$ & $-0.299 \pm 0.215$ & $-0.0502 \pm 0.0549$ & $-0.323 \pm 0.201$ & $-0.1533 \pm 0.0631$ \\
$b_{\Pi^{[2]}_\parallel}$ & $-0.0344 \pm 0.0508$ & $-0.0168 \pm 0.0250$ & $-0.0223 \pm 0.0537$ & $+0.0057 \pm 0.0284$ \\
$b_1^q$ & $+3.2950 \pm 0.0352$ & $+3.3154 \pm 0.0157$ & $+3.3095 \pm 0.0221$ & $+3.3137 \pm 0.0146$ \\
$b_2^q$ & $+2.218 \pm 0.875$ & $+2.717 \pm 0.780$ & $+1.60 \pm 1.29$ & $+2.084 \pm 0.738$ \\
$b_{\mathcal{G}_2}^q$ & $-0.84 \pm 1.29$ & $-0.89 \pm 1.14$ & $+0.54 \pm 1.23$ & $+0.45 \pm 1.11$ \\
\hline
\multicolumn{5}{@{}l}{\textbf{Analytically marginalized parameters for the power spectrum}} \\
$b_{\Gamma_3}$ & $-0.089 \pm 0.575$ & $-0.222 \pm 0.583$ & $-0.406 \pm 0.623$ & $-0.449 \pm 0.611$ \\
$c_{s,0}$ & $+0.1312 \pm 0.0174$ & $+0.0365 \pm 0.0176$ & $+0.0278 \pm 0.0106$ & $+0.0114 \pm 0.0105$ \\
$c_{s,2}$ & $-0.0155 \pm 0.0255$ & $+0.0525 \pm 0.0256$ & $-0.1266 \pm 0.0240$ & $-0.0821 \pm 0.0237$ \\
$c_{s,4}$ & $-0.3216 \pm 0.0198$ & $-0.3225 \pm 0.0198$ & $-0.5633 \pm 0.0212$ & $-0.5690 \pm 0.0211$ \\
$b_{\Pi^{[3]}_\parallel}$ & $+0.240 \pm 0.404$ & $+0.119 \pm 0.410$ & $+0.264 \pm 0.438$ & $+0.303 \pm 0.428$ \\
$b_{\delta\Pi^{[2]}_\parallel}$ & $-1.382 \pm 0.522$ & $-1.229 \pm 0.530$ & $-2.420 \pm 0.569$ & $-2.777 \pm 0.556$ \\
$b_{K\Pi^{[2]}_\parallel}$ & $-0.65 \pm 1.33$ & $-0.79 \pm 1.35$ & $-1.58 \pm 1.44$ & $-1.80 \pm 1.42$ \\
$b_{\eta\Pi^{[2]}_\parallel}$ & $-2.794 \pm 0.298$ & $-3.378 \pm 0.302$ & $-6.331 \pm 0.361$ & $-7.361 \pm 0.354$ \\
$P_{\rm shot}$ & $-0.4275 \pm 0.0528$ & $-0.1961 \pm 0.0529$ & $-0.1207 \pm 0.0236$ & $-0.1031 \pm 0.0239$ \\
$a_0$ & $+1.040 \pm 0.226$ & $+0.827 \pm 0.228$ & $+0.752 \pm 0.167$ & $+0.829 \pm 0.168$ \\
$a_2$ & $-3.977 \pm 0.653$ & $-3.530 \pm 0.650$ & $-8.299 \pm 0.756$ & $-8.101 \pm 0.756$ \\
$b_{\Gamma_3}^q$ & $+0.478 \pm 0.137$ & $-0.461 \pm 0.139$ & $-2.223 \pm 0.146$ & $-3.465 \pm 0.143$ \\
$c_{s,0}^q$ & $+1.249 \pm 0.400$ & $+1.947 \pm 0.403$ & $-0.073 \pm 0.301$ & $-0.311 \pm 0.300$ \\
$c_{s,2}^q$ & $+7.224 \pm 0.627$ & $+3.927 \pm 0.631$ & $+7.572 \pm 0.602$ & $+6.975 \pm 0.600$ \\
$c_{s,4}^q$ & $-2.259 \pm 0.673$ & $-1.455 \pm 0.676$ & $-6.372 \pm 0.702$ & $-6.146 \pm 0.701$ \\
$b_{4,\times}$ & $+5.33 \pm 1.88$ & $+6.84 \pm 1.87$ & $+12.39 \pm 1.78$ & $+12.39 \pm 1.78$ \\
$a_4$ & $+8.12 \pm 1.08$ & $+7.10 \pm 1.08$ & $+27.82 \pm 1.53$ & $+26.68 \pm 1.53$ \\
$a_{4,\times}$ & $-0.001 \pm 0.742$ & $-0.001 \pm 0.742$ & $-0.000 \pm 0.742$ & $-0.000 \pm 0.742$ \\
\hline
\multicolumn{5}{@{}l}{\textbf{Linear bispectrum amplitudes}} \\
$B_{\rm shot}$ & -- & $-0.0143 \pm 0.0145$ & -- & $-0.0131 \pm 0.0132$ \\
$C_{\rm shot}$ & -- & $+0.0992 \pm 0.0969$ & -- & $+0.0974 \pm 0.0968$ \\
$A_{\rm shot}$ & -- & $+0.0100 \pm 0.0100$ & -- & $+0.0100 \pm 0.0100$ \\
$c_0^B$ & -- & $-0.668 \pm 0.604$ & -- & $-0.054 \pm 0.724$ \\
$c_2^B$ & -- & $-0.077 \pm 0.972$ & -- & $-0.01 \pm 1.00$ \\
$c_4^B$ & -- & $+4.52 \pm 2.39$ & -- & $+3.60 \pm 4.89$ \\
\hline
\multicolumn{5}{@{}l}{\textbf{Goodness of fit}} \\
$\chi^2_P / N_P$ & 926.8 / 1772 & 931.2 / 1772 & 1069.7 / 1772 & 1065.2 / 1772 \\
$\chi^2_B / N_B$ & -- & 129.5 / 230 & -- & 133.0 / 230 \\
\hline
\end{tabular}
}
    \caption{
        \textbf{Free-cosmology fits to the mean data vector of Models 1 and 3.} Summary of the 68\% C.L. 1D marginalized constraints for the chains  illustrated in Fig.~\ref{fig:freecosmo_compare_M1_M3}.
    }
    \label{tab:tab_meandv_freecosmo}
\end{table}

\section{Mean power spectrum data vector}\label{app:mean_power_spectrum}

\begin{figure}
    \centering
    \includegraphics[width=0.49\linewidth]{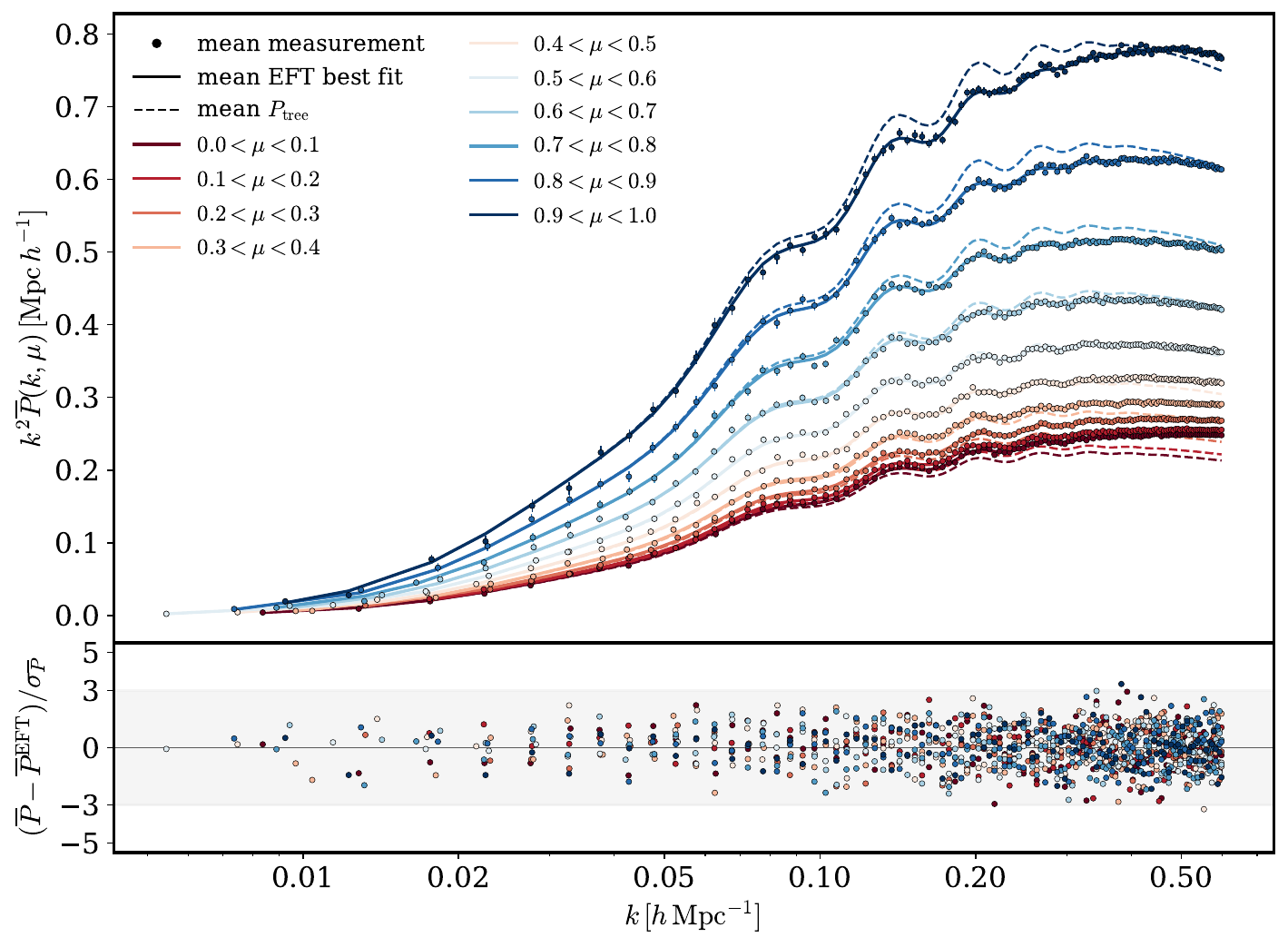}\hfill
    \includegraphics[width=0.49\linewidth]{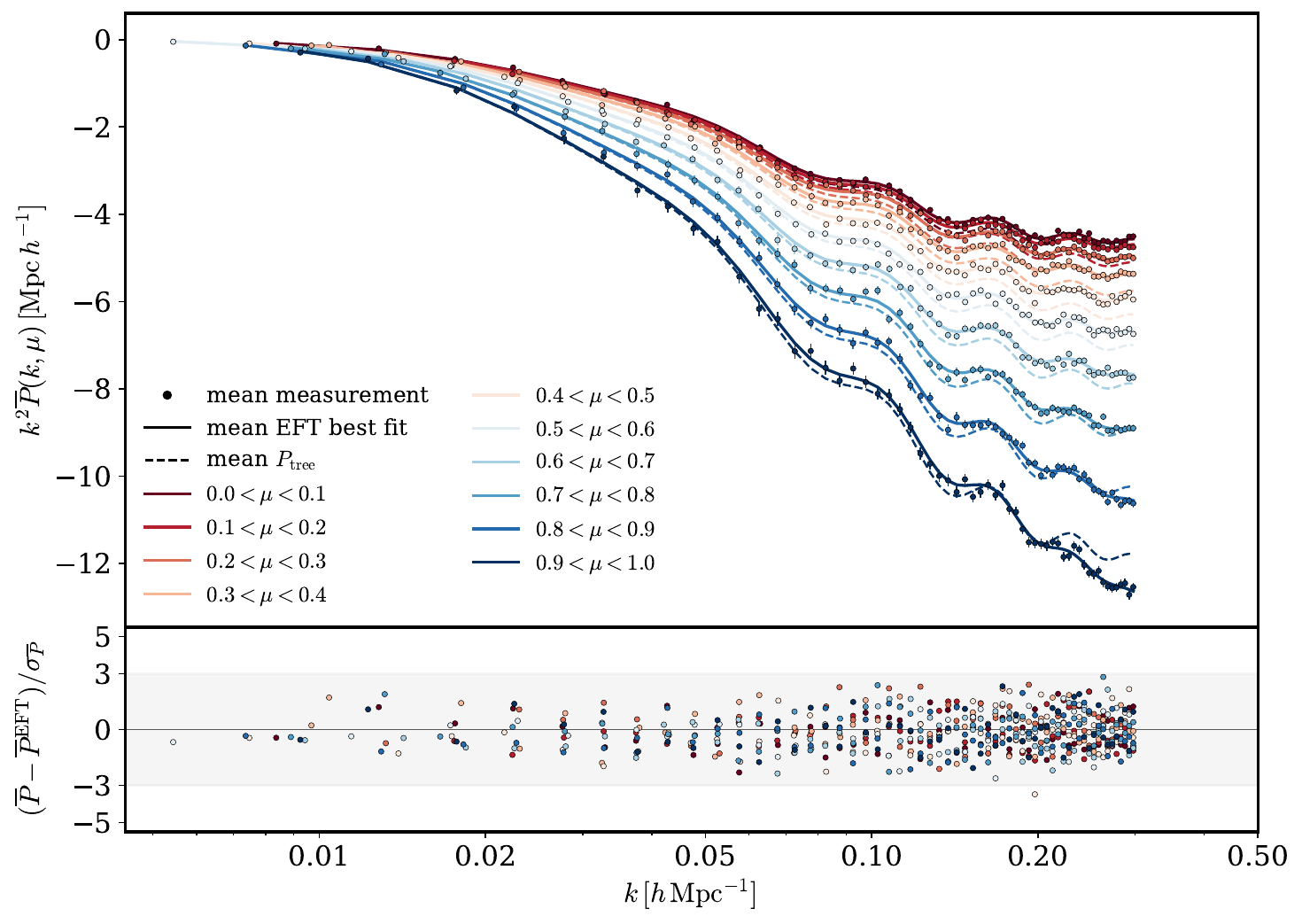}
    \vspace{-0.1in}
    \caption{\textbf{Mean power spectrum measurement.} Mean EFT fits to the \Lya auto-correlation (left panel) and \Lya-quasar cross-correlation (right panel) function in ten angular bins $\mu$, with a maximum scale of $\kmax=0.6\hMpcinv$ for the simulations of model 1. The best-fit parameters are shown in Tab.~\ref{tab:tab_meandv_freecosmo}. The fits are obtained jointly with the bispectrum, discussed in Fig.~\ref{fig:freecosmo_B_mean_equilater_vs_squeezed}. We compare the best-fit EFT model (solid line) to tree-level (dashed line) power spectrum and find a large deviation beyond $k>0.1\hMpcinv$. The error bars are obtained assuming a diagonal Gaussian covariance based on the number of expected Fourier modes per bin. The residuals between the model and the data are shown in the bottom panel with the error on the mean. To guide the eye we include a $3\sigma$ band in gray. }
    \label{fig:freecosmo_PB_mean}
\end{figure}

To build more intuition about the quality of the fits we show in Fig.~\ref{fig:freecosmo_PB_mean} the resulting best-fit power spectra for $FF$ (left) and $FQ$ (right) in the ten angular bins used for the analysis for Fig.~\ref{fig:freecosmo_compare_M1_M3}. We compare the EFT predictions (solid lines) to the tree-level power spectrum model (dashed lines) which is evaluated at the best-fit linear bias parameters and find strong deviations for all angular wedges beyond $k>0.1,\,(0.3)\hMpcinv$ exceeding the 10 (30) \% level. The bottom panel for both figures shows the residuals which are within the $3\sigma$ gray error band which we included to guide the eye. Note that the error bars are the errors on the mean which is a challenging test for the methodology.

\end{document}